\documentclass[draftclsnofoot,onecolumn, 12pt]{IEEEtran}
\usepackage{graphicx, pstool}
\usepackage{amsmath}
\usepackage{amsfonts}
\usepackage{amssymb}
\usepackage{booktabs}
\usepackage{algorithm,algorithmic}
\usepackage{float}
\usepackage{url}

\newtheorem{theorem}{Theorem}
\newtheorem{remark}{Remark}
\newtheorem{lemma}{Lemma}

\begin{document}
\title{Capacity of the Gaussian  Two-Hop Full-Duplex Relay Channel with   Residual  Self-Interference
}
\author{\normalsize Nikola Zlatanov, Erik Sippel, Vahid Jamali,   and   Robert
Schober
\thanks{This work   was accepted  in part for presentation at  IEEE Globecom 2016  \cite{C_FD_SI_conf}.}
\thanks{N. Zlatanov is  with the Department of Electrical and Computer Systems Engineering, Monash University, Melbourne, VIC 3800, Australia (e-mail: nikola.zlatanov@monash.edu).}
\thanks{ E. Sippel, V. Jamali, and R. Schober are with the Friedrich-Alexander University of Erlangen-N\"urnberg,
Institute for Digital Communications, D-91058 Erlangen, Germany
(e-mails: (erik.sippel@fau.de, vahid.jamali@fau.de  robert.schober@fau.de).
}
\vspace{-14mm}
}

\maketitle

\begin{abstract}
In this paper, we investigate the capacity of the Gaussian two-hop  full-duplex (FD) relay channel  with  residual   self-interference. This channel is comprised of a source,  an FD relay, and a destination, where a direct source-destination  link does not exist  and  the FD relay is impaired by  residual  self-interference.  We adopt the worst-case  linear self-interference model with respect to the channel capacity, and model the  residual  self-interference as a Gaussian random variable whose  variance    depends on the  amplitude of the transmit symbol  of the relay. For this channel, we derive the capacity and propose an explicit capacity-achieving coding scheme. Thereby, we show that the optimal input distribution at  the source is Gaussian and its variance  depends on the  amplitude  of the transmit symbol of the relay.   On the other hand,  the optimal input distribution at the relay is discrete or Gaussian, where the latter case occurs  only   when the relay-destination link is the bottleneck link. The derived capacity  converges to the capacity of the two-hop ideal FD relay channel  without self-interference and to the capacity of the two-hop half-duplex (HD)  relay channel in the limiting cases when the  residual  self-interference is zero and infinite, respectively.  Our numerical results show that significant performance gains are achieved with the proposed capacity-achieving coding scheme  compared to the achievable rates of conventional HD relaying  and/or conventional FD relaying. 
\end{abstract}

\section{Introduction}
In wireless communications, relays are employed in order to increase the data rate between a source and a destination. The resulting  three-node   channel is known as the relay channel \cite{cover}. If the distance between the source and the destination is very large or there is  heavy blockage, then the   relay channel can be modeled without a source-destination link, which leads to the so called two-hop relay channel. For the relay channel, there are two different modes of  operation for the relay,  namely, the full-duplex (FD) mode and  the half-duplex (HD) mode. In the FD mode, the relay transmits and receives at the same time and in the same frequency band. As a result,   FD relays are impaired by self-interference, which is the interference caused by the relay's transmit signal to the relay's received signal. Latest advances in hardware design have shown that  the self-interference of an FD node can be suppressed significantly, see  
\nocite{5089955, Choi:2010, 5961159, 5985554, Jain_2011, 6177689,    6280258, 6353396, Bharadia:2013:FDR:2486001.2486033, 6542771, 6523998, 6702851,6736751, 6656015, 6782415, 6832592, 6862895, 6832471,  6832464,   6832439,  7105647, 7024120,      7051286,    7390828,  6736751, 7182305} \cite{5089955}-\cite{7182305}, which has led to an enormous interest in FD communication.  For example, \cite{Bharadia:2013:FDR:2486001.2486033} reported that self-interference suppression of 110 dB   is possible in certain scenarios. 
 On the other hand, in the HD mode, the relay transmits and receives  in the same frequency band but in different time slots or in the same time slot but in different frequency bands. As a result,   HD relays completely   avoid  self-interference. However, since an HD relay transmits and receives only in half of the time/frequency resources compared to an FD relay, the achievable rate of the two-hop HD relay channel  may be  significantly lower than that of the  two-hop FD relay channel.  

Information-theoretic  analyses of the capacity of the two-hop HD relay channel  were provided in \cite{kramer2004models}, \cite{zlatanov2014capacity-globecom}. Thereby, it was shown that the capacity of the two-hop HD relay channel  is achieved when the HD relay switches between reception and transmission in a symbol-by-symbol manner and not in a codeword-by-codeword manner, as is done in conventional HD relaying \cite{1435648}. Moreover,
in order to achieve the capacity, the HD relay has to encode information into the silent symbol   created when the relay receives \cite{zlatanov2014capacity-globecom}. 
For the Gaussian two-hop HD relay channel without fading, it was shown  in \cite{zlatanov2014capacity-globecom} that the optimal input distribution at the relay is discrete and includes the zero (i.e., silent) symbol. On the other hand,  the source transmits using a Gaussian input distribution  when the relay transmits the zero (i.e., silent) symbol and is silent otherwise. 

 The capacity of the Gaussian two-hop FD relay channel with  ideal FD relaying  without residual self-interference was derived in   \cite{cover}. However, in practice, canceling the residual self-interference completely is not possible due to  limitations in  channel estimation precision and imperfections in the transceiver design \cite{6832464}.  
As a result, the residual self-interference has to be taken into  account when investigating the capacity of the two-hop FD relay channel.   
Despite the considerable body of work on  FD relaying, see  e.g. \cite{5961159, 5985554, 6280258, 6862895,  7390828},  the capacity of the two-hop FD relay channel with residual self-interference has  not been explicitly characterized yet. As a result, for this channel,  only achievable rates are known  which are strictly smaller than the capacity.
 Therefore, in this paper, we study the capacity of the  two-hop FD relay channel with residual self-interference for the case when the source-relay and relay-destination links are  additive white Gaussian noise (AWGN) channels.

In general, the statistics  of the residual self-interference   depend  on the employed hardware configuration and  the adopted self-interference suppression schemes. As a result, different  hardware configurations and  different self-interference suppression schemes may lead to different  statistical properties of the residual self-interference, and thereby, to different capacities for the considered relay channel. An upper bound on the capacity of the two-hop FD relay channel with residual self-interference is given in in \cite{cover} and is obtained by assuming zero residual  self-interference. Hence, the objective of this paper is to derive
  a lower bound on the  capacity of this channel valid for any    linear residual self-interference model. To this end, we consider  the worst-case  linear self-interference model with respect to the capacity, and thereby, we obtain the desired lower bound on the capacity for any other type  of   linear residual self-interference.
 For the  worst-case, the linear residual self-interference is    modeled   as a conditionally  Gaussian distributed random variable (RV) whose variance   depends on the amplitude of the symbol transmitted by the relay.

For this relay channel, we
derive the corresponding capacity and propose an explicit coding scheme which achieves the capacity. We show that the FD relay has to operate in the decode-and-forward (DF) mode to achieve the capacity, i.e., it has to decode each  codeword received from the source  and then transmit the decoded information to the destination in the next time slot, while simultaneously receiving. Moreover, we show that   the optimal input distribution at the relay is 
discrete or Gaussian, where the latter case occurs only when the relay-destination link is the
bottleneck link. On the other hand,   the capacity-achieving input distribution at the source is Gaussian and its variance
depends on the amplitude of the symbol transmitted by the relay, i.e., the average power of the source's transmit symbol  depends on the amplitude of the relay's transmit symbol. In particular, the smaller the amplitude of
the relay's transmit symbol   is, the higher the average power of the source's transmit symbol 
should be since, in that case, the residual self-interference is small with high probability. On the other hand,  if the amplitude of the relay's transmit symbol is very large and exceeds some   threshold,  the chance for very strong residual  self-interference is high   and the source should remain  silent
and  conserve its energy for other symbol intervals  with weaker   residual  self-interference. We show that the derived capacity  converges to the capacity of the two-hop ideal FD relay channel  without self-interference \cite{cover} and to the capacity of the two-hop  HD   relay channel \cite{zlatanov2014capacity-globecom} in the limiting cases when the residual self-interference is zero and infinite, respectively. Our   numerical results reveal that significant performance gains are achieved with the proposed capacity-achieving coding scheme  compared to the achievable rates of conventional HD relaying  and/or conventional FD relaying. 

This paper is organized as follows. In Section~\ref{Sec2}, we present the models for the  channel and the residual self-interference. In Section~\ref{Sec3}, we present the capacity of the considered channel and propose an explicit capacity-achieving coding scheme.   Numerical examples are provided in   Section~\ref{Sec-Num}, and Section~\ref{con} concludes the paper.

\section{System  Model}\label{Sec2}

In the following, we introduce the models  for the two-hop FD relay channel   and the residual self-interference.

\subsection{Channel Model}
We assume a two-hop FD relay channel comprised of a source, an FD relay, and a destination, where a direct source-destination link does not exist. We assume that the source-relay and the relay-destination links are  AWGN channels, and that the FD relay is impaired by residual self-interference. In symbol interval $i$, let $X_S[i]$ and $X_R[i]$ denote RVs which model the  transmit  symbols at the source and the relay, respectively, let $\hat Y_R[i]$ and $\hat Y_D[i]$ denote     RVs  which model the received symbols at the relay and the destination, respectively,    and let $\hat N_R[i]$ and $\hat N_D[i]$ denote RVs which model the AWGNs  at the relay and the destination, respectively. We assume that $\hat N_R[i]\sim\mathcal{N}(0,\hat\sigma_R^2)$ and $\hat N_D[i]\sim\mathcal{N}(0,\hat \sigma_D^2)$, $\forall i$, where $\mathcal{N}(\mu, \sigma^2)$ denotes a Gaussian distribution with mean $\mu$ and variance $\sigma^2$.  Moreover, let $h_{SR}$ and  $h_{RD}$  denote the channel gains of the source-relay and relay-destination channels, respectively, which are assumed to be constant during all symbol intervals, i.e., fading\footnote{As  customary for capacity analysis, see e.g.  \cite{cover2012elements}, as a first step  we do not consider fading and assume  real-valued channel inputs and outputs.  The generalization to a complex-valued signal model   is relatively straightforward \cite{TSE05}. On the other hand, the generalization to the case of fading is considerably more involved. For example, considering the achievability  scheme for HD relays in \cite{BA-relaying-adaptive-rate}, we expect that  when fading is present, both   HD and FD relays have to perform buffering in order to achieve the capacity. However, the corresponding detailed analysis is beyond the scope of this paper and   presents an interesting topic for future research.} is not considered. In addition, let  $\hat I[i]$ denote  the RV which models the residual  self-interference  at the FD relay  that remains  in symbol interval $i$ after analog and digital self-interference cancelation.

 Using the notations defined above, the input-output relations describing the considered relay channel in symbol interval $i$ are given as
\begin{align}
\hat Y_R[i]&=h_{SR}   X_S[i]+ \hat I[i]+\hat N_R[i]\label{r1}\\
\hat Y_D[i]&=h_{RD}    X_R[i]+\hat N_D[i].\label{r2}
\end{align}
Furthermore, an  average ``per-node''  power constraint  is assumed, i.e.,
\begin{align}
E\{X_{\beta}^2[i]\} &= \lim\limits_{n\to \infty}\frac{1}{n}\sum_{k=1}^{n}X_{\beta}^2[k]  \le P_{\beta},\quad\beta\in\{S,R\},  \label{con3a}
\end{align}
where $E\{\cdot\}$ denotes statistical expectation, and $P_S$ and $P_R$ are the average power constraints at the source and the relay, respectively.  
 
\subsection{Residual Self-Interference  Model}\label{sec_2-2}

Assuming narrow-band signals  such that the channels can be modelled as   frequency flat, a general model for the residual self-interference  at the FD relay in symbol interval $i$, $\hat I[i]$, is given  by \cite{Bharadia:2013:FDR:2486001.2486033} 
\begin{align}\label{r1_eq_1}
\hat I[i]=\sum_{m=1}^M   h_{RR,m}[i] \big(X_R[i]\big)^m,
\end{align}
where $M\leq \infty$ is an integer  and $ h_{RR,m}[i]$ is the residual  self-interference channel between the   transmitter-end and the receiver-end of the FD relay through  which  symbol $\big(X_R[i]\big)^m$ arrives at the   receiver-end. Moreover,  for $m=1$, $\big(X_R[i]\big)^m$  is the linear component of the residual  self-interference, and  for $m\geq 2$, $\big(X_R[i]\big)^m$  is a nonlinear component  of the residual  self-interference. As shown in \cite{Bharadia:2013:FDR:2486001.2486033}, only the terms  for  which $m$ is odd in   (\ref{r1_eq_1}) carry non-negligible energy while the remaining  terms for  which $m$ is even can be ignored. Moreover,  as observed in \cite{Bharadia:2013:FDR:2486001.2486033}, the higher order terms  in   (\ref{r1_eq_1}) carry significantly less energy than the lower order terms, i.e., the term for $m=5$ carries significantly less energy than the term for   $m=3$, and the   term  for $m=3$ carries significantly less energy than the term for   $m=1$.   As a result, in this paper, we adopt the first order approximation of the residual  self-interference  in (\ref{r1_eq_1}), i.e.,  $\hat I[i]$ is modeled as
\begin{align}\label{r1_eq_2}
\hat I[i]\approx    h_{RR}[i] X_R[i],
\end{align}
where $ h_{RR}[i]=  h_{RR,1}[i]$ is used for simplicity of notation. Obviously, the residual  self-interference model in  (\ref{r1_eq_2})  takes into account only the linear component of the residual  self-interference and assumes that the nonlinear components can be neglected. Such a linear model for the residual self-interference is particularly justified for  relays with low average transmit powers \cite{6353396}.

The residual  self-interference channel gain in (\ref{r1_eq_2}), $  h_{RR}[i]$, is time-varying   even  when fading is not present, see e.g. \cite{5985554, 6353396, 6782415,  6832439}. The variations of the  residual  self-interference channel gain, $  h_{RR}[i]$, are due to the cumulative effects of various   distortions originating from noise, carrier frequency offset, oscillator phase noise, analog-to-digital/digital-to-analog (AD/DA) conversion imperfections, I/Q imbalance,    imperfect  channel estimation, etc., see \cite{5985554, 6353396, 6782415,  6832439}. These  distortions\footnote{We note that similar  distortions  are also present  in the source-relay and relay-destination channels. However, due to the large distance between  transmitter  and   receiver, the impact of these distortions  on the channel gains $h_{SR}$ and  $h_{RD}$ is negligible.} have a significant impact on the residual  self-interference channel gain due to the very small distance between the transmitter-end  and the receiver-end of the   self-interference channel.  Moreover,  the  variations of the residual   self-interference channel gain, $ h_{RR}[i]$, are random and thereby cannot be accurately estimated at the FD node \cite{5985554, 6353396, 6782415,  6832439}. The statistical properties of these   variations  are dependent on the employed hardware configuration and the adopted  self-interference suppression schemes. In \cite{6353396},   $ h_{RR}[i]$ is assumed to be constant during one codeword comprised of many symbols. Thereby, the residual  self-interference model in \cite{6353396} models only the long-term, i.e., codeword-by-codeword,  statistical properties of the residual  self-interference. However,   the symbol-by-symbol variations of  $  h_{RR}[i]$  are not captured by the model in \cite{6353396}  since they are averaged out. Nevertheless, for a meaningful  information-theoretical analysis, the statistics of the symbol-by-symbol   variations of $ h_{RR}[i]$ are needed. On the other hand, the statistics of the variations of   $ h_{RR}[i]$ affect the capacity of the considered relay channel. In this paper, we derive the capacity of the considered relay channel for the worst-case  linear residual  self-interference model, which yields a lower bound for the capacity for any other   linear residual  self-interference model.

To derive the worst-case linear residual self-interference model in terms of capacity, we insert   (\ref{r1_eq_2}) into (\ref{r1}), and obtain the  received symbol at the relay in symbol interval $i$ as
\begin{align}
\hat Y_R[i] =     h_{SR}   X_S[i]    +    h_{RR}  X_R[i] + \hat N_R[i]. \label{r1v1a} 
\end{align}
Now, since  in general $h_{RR}[i]$ can  have a non-zero mean, without loss of generality, we can write $h_{RR}[i]$  as
\vspace*{-1mm}
\begin{align}
h_{RR}[i] &= \bar h_{RR} +   \hat h_{RR}[i] ,  \label{eq_eq_x-h1}
\end{align}
where $\bar h_{RR}$ is the mean of $h_{RR}[i]$, i.e., $\bar h_{RR}=E\{h_{RR}[i]\}$  and $\hat h_{RR}[i]=h_{RR}[i]-\bar h_{RR}$  is the remaining zero-mean random component of $h_{RR}[i]$.
Inserting (\ref{eq_eq_x-h1})   into (\ref{r1v1a}),   we obtain the  received symbol at the relay in symbol interval $i$ as
\begin{align}
\hat Y_R[i]&=     h_{SR}   X_S[i]    + \bar h_{RR}  X_R[i]+    \hat h_{RR}[i]  X_R[i] + \hat N_R[i]. \label{r1v1} 
\end{align}
Given sufficient time, the relay can   estimate any mean in its received symbols arbitrarily accurately, see \cite{841172}. Thereby, given sufficient time, the relay can estimate  the deterministic component of the residual  self-interference channel gain $\bar h_{RR}$. Moreover,  since $X_R[i]$ models the desired  transmit symbol at the relay, and since the relay knows which symbol it wants to transmits,   the outcome of $X_R[i]$, denoted by $x_R[i]$, is known in each symbol interval $i$. As a result, the relay knows  $\bar h_{RR} X_R[i]$ and thereby it can subtract   $\bar h_{RR} X_R[i]$ from the received symbol $\hat Y_R[i]$ in (\ref{r1v1}).  Consequently,  we obtain a new received symbol at the relay in symbol interval $i$, denoted by $\tilde Y_R[i]$, as
 \begin{align}
\tilde Y_R[i]&=   h_{SR} X_S[i]      +   \hat h_{RR}[i]   X_R[i] + \hat  N_R[i] .   
\label{r1v3a} 
\end{align}
Now, assuming that the relay transmits   symbol $x_R[i]$ in symbol interval $i$, i.e.,   $X_R[i]=x_R[i]$, from (\ref{r1v3a}) we conclude that the relay ``sees''  the following additive impairment in symbol interval $i$
\begin{align}\label{eq_r1-eq3}
    \hat h_{RR}[i]  x_R[i] + \hat N_R[i]. 
\end{align}
Consequently, from  (\ref{eq_r1-eq3}), we conclude that the worst-case scenario  with respect to the capacity is if (\ref{eq_r1-eq3}) is zero-mean  independent and identically   distributed (i.i.d.)   Gaussian RV\footnote{This is because a  Gaussian RV has the highest uncertainty (i.e., entropy) among all possible RVs for a given second moment \cite{cover2012elements}.}, which is possible only if $ \hat h_{RR}[i] $ is a zero-mean i.i.d. Gaussian RV. 
Hence, modeling the  linear residual self-interference channel gain, $  h_{RR}[i]$, as an   i.i.d. Gaussian RV  constitutes the worst-case linear  residual self-interference model,  and thereby, leads to a lower bound on the capacity for any other distribution of $ h_{RR}[i]$.    

Considering the developed worst-case linear residual self-interference model, in the rest of this paper, we assume $ \hat h_{RR}[i]\sim\mathcal{N}(0,\hat\alpha) $, where   $\hat \alpha$ is the variance of $\hat h_{RR}[i]$. Since the average power  of the linear residual self-interference at the relay is $\hat \alpha E\{X_R^2[i]\}$,   $\hat \alpha$ can   be interpreted as a   self-interference amplification factor, i.e.,  $1/\hat \alpha$ is a  self-interference suppression factor.

\subsection{Simplified Input-Output Relations for the Considered Relay Channel}

 To simplify the input-output relation in (\ref{r1v3a}), we divide the  received symbol $\tilde Y_R[i]$  by   $h_{SR}$ and thereby  obtain a new received symbol at the relay in symbol interval $i$, denoted by $Y_R[i]$, and given by
 \begin{align}
Y_R[i]&=    X_S[i]      + \frac{ \hat h_{RR}[i]  }{h_{SR}} X_R[i] +\frac{\hat  N_R[i]}{h_{SR}}   = X_S[i]      +   I[i] +   N_R[i], 
\label{r1v3} 
\end{align}
where
\begin{align}\label{eq_asd}
 I[i]= \frac{\hat h_{RR}[i]  }{h_{SR}} X_R[i] 
\end{align}
is the normalized residual self-interference at the relay and $
N_R[i]=  \hat  N_R[i] /h_{SR}$
is the normalized noise at the relay  distributed according to $N_R[i]\sim\mathcal{N}(0, \sigma_R^2)$, where $\sigma_R^2= \hat\sigma_R^2/h_{SR}^2$. The normalized  residual self-interference, $ I[i]$,  is  dependent on the transmit symbol at the relay, $X_R[i]$, and, conditioned on $X_R[i]$, it has the same type of  distribution as  the  random component  of the self-interference channel gain, $\hat h_{RR}[i]$, i.e., an i.i.d. Gaussian distribution. Let     $\alpha$ be defined as $\alpha=\hat \alpha/h_{SR}^2$,  which can be interpreted as   the normalized self-interference amplification factor.  Using $\alpha$ and    assuming that  the transmit symbol at the relay in symbol interval $i$ is  $X_R[i]=x_R[i]$,   the distribution of the normalized  residual self-interference, $ I[i]$, can be written as
\begin{align}\label{eq_sdxx2}
 I[i]\sim\mathcal{N}(0,  \alpha  x_R^2[i]),\; \textrm{ if } \; X_R[i]=x_R[i].
\end{align}

To obtain also a normalized received symbol at the destination,  we normalize $\hat Y_D[i]$ in (\ref{r2}) by $h_{RD}$, which yields
\begin{align}
Y_D[i]&= X_R[i]+N_D[i].\label{r2a}
\end{align}
In (\ref{r2a}), $N_D[i]$ is the normalized noise power at the destination distributed as $N_D[i]  =  \mathcal{N}\left(0, \sigma_D^2 \right) $,
where    $ \sigma_D^2 = \hat \sigma_D^2/h^2_{RD} $.

Now, instead of deriving the capacity of the considered relay channel using the input-output relations in (\ref{r1}) and (\ref{r2}), we can derive the capacity using an equivalent relay channel defined by the input-output relations in (\ref{r1v3}) and (\ref{r2a}), respectively, where, in symbol interval $i$, $X_S[i]$ and $X_R[i]$ are the inputs at source and relay, respectively,  $Y_R[i]$ and $Y_D[i]$ are the outputs at   relay and destination, respectively,   $N_R[i]$ and $N_D[i]$ are   AWGNs with  variances $\sigma_R^2= \hat \sigma_R^2/h_{SR}^2$ and $\sigma_D^2=\hat \sigma_D^2/h_{RD}^2$, respectively, and   $ I[i]$ is the residual self-interference with conditional distribution given by (\ref{eq_sdxx2}), which is a function of the  normalized self-interference amplification factor  $ \alpha$.

\section{Capacity}\label{Sec3}

In this section, we study the capacity of the considered Gaussian  two-hop FD relay channel with residual self-interference.

\subsection{Derivation of the Capacity}
To derive the capacity of the considered relay channel, we first assume that  RVs  $X_S$ and $X_R$, which model the transmit  symbols at source and relay for any symbol interval $i$,    take   values $x_S$ and $x_R$ from sets $\mathcal{X}_S$ and $\mathcal{X}_R$, respectively.  Now, since the considered relay channel belongs to the class of     memoryless degraded relay channels defined in \cite{cover}, its capacity is given by  \cite[Theorem 1]{cover} 
\begin{align}\label{con2}
C=\max_{p(x_S,x_R)\in\mathcal{P}}&~\min\big\{I(X_S;Y_R|X_R),I(X_R;Y_D)\big\}
\nonumber\\
\textrm{Subject to}  \textrm{ C1: }& E\{X_S^2\}\leq P_S\nonumber\\
\textrm{ C2: }& E\{X_R^2\}\leq P_R,
\end{align}
where $\mathcal{P}$ is a set which contains all valid distributions.
In order to obtain the capacity  in (\ref{con2}), we need to find the optimal joint input distribution, $p(x_S,x_R)$, which maximizes  the $\min\{\cdot\}$ function in (\ref{con2}) and satisfies constraints C1 and C2. To this end, note that  $p(x_S,x_R)$ can be written equivalently  as 
$
p(x_S,x_R)=p(x_S|x_R)p(x_R).
$
Using this relation, we can represent the maximization in (\ref{con2}) equivalently as two nested maximizations, one with respect to $p(x_S|x_R)$ for a fixed $p(x_R)$, and the other one with respect to $p(x_R)$. Thereby, we can write the capacity in (\ref{con2}) equivalently as
\begin{align} 
C=\max_{p(x_R)\in\mathcal{P}}~\max_{p(x_S|x_R)\in\mathcal{P}} ~ & \mathrm{min}\big\{I(X_S;Y_R|X_R),I(X_R;Y_D)\big\}\nonumber\\
\textrm{Subject to}  \textrm{ C1: }& E\{X_S^2\}\leq P_S\nonumber\\
\textrm{ C2: }& E\{X_R^2\}\leq P_R.\label{con2a}
\end{align}
Now, since in the $\min\{\cdot\}$ function in (\ref{con2a}) only $I(X_S;Y_R|X_R)$ is dependent on the distribution $p(x_S|x_R)$, whereas $I(X_R;Y_D)$ does not depend on  $p(x_S|x_R)$, we can write the capacity expression in   (\ref{con2a}) equivalently as
\begin{align} 
C=\max_{p(x_R)\in\mathcal{P}} &~ \mathrm{min}\left\{\max_{p(x_S|x_R)\in\mathcal{P}}I(X_S;Y_R|X_R),I(X_R;Y_D)\right\}\nonumber\\
\textrm{Subject to}  \textrm{ C1: }& E\{X_S^2\}\leq P_S\nonumber\\
\textrm{ C2: }& E\{X_R^2\}\leq P_R. \label{con2b}
\end{align}
Hence, to obtain the capacity of the considered relay channel, we first need to find the conditional input distribution at the source, $p(x_S|x_R)$, which maximizes $I(X_S;Y_R|X_R)$ such that constraint C1 holds. Next, we need to find the optimal input distribution at the relay, $p(x_R)$, which maximizes the $\min\{\cdot\}$ expression in (\ref{con2b}) such that constraints C1 and C2 hold.  

\subsection{Optimal Input Distribution at the  Source $p^*(x_S|x_R)$}

The optimal input distribution at the source  which achieves the capacity in (\ref{con2b}), denoted by $p^*(x_S|x_R)$, is given in the following theorem.

\begin{theorem}\label{Theo1}
 The optimal    input distribution at the source  $p^*(x_S|x_R)$, which achieves the capacity of the considered relay channel  in (\ref{con2b}), is the zero-mean Gaussian distribution with  variance $P_S(x_R)$ given by
 \begin{align}\label{P1}
 P_S(x_R)=\alpha\max\{0,x_{\rm th}^2- x_R^2\},
 \end{align}
 where  $x_{\rm th}$ is a positive threshold constant found as follows. 
If  $p(x_R)$ is a discrete distribution,  $x_{\rm th}$ is found  as the solution of the following identity
	\begin{gather}
	\sum_{x_R\in \mathcal{X}_R} \alpha  \max\{0,x_{\rm th}^2-   x_R^2\} p(x_R) =  P_S,  \label{36a}
	\end{gather}
 and the corresponding $\max\limits_{p(x_S|x_R)\in\mathcal{P}}  I(X_S;Y_R|X_R)$ is obtained as
\begin{align}\label{eq_1-dis}
    \max_{p(x_S|x_R)\in\mathcal{P}}   I(X_S;Y_R|X_R) 
 = 
\sum\limits_{x_R\in \mathcal{X}_R} \frac{1}{2} \log_2\left(1+\frac{\alpha \max\{0,\;x_{\rm th}^2- x_R^2\}}{\sigma_R^2+\alpha x_R^2}\right) p(x_R) .   
\end{align}
Otherwise, if  $p(x_R)$ is a continuous distribution, the sums in  (\ref{36a}) and (\ref{eq_1-dis}) have to be replaced by   integrals.
\end{theorem}

\begin{IEEEproof}
Please refer to  Appendix A.
\end{IEEEproof}

From Theorem~\ref{Theo1}, we can see that the source should perform power allocation in a symbol-by-symbol manner. In particular, the average power of the source's transmit symbols, $P_S(x_R)$, given by (\ref{P1}),  depends on the amplitude of the transmit symbol at the relay, $|x_R|$. The lower the amplitude of the transmit symbol of the relay is,  the higher the average power of the source's transmit symbols should be since, in that case, there is a high probability for weak residual self-interference. Conversely, the higher the amplitude of the transmit symbol of the relay is,  the lower  the average power of the source's transmit symbols should be since, in that case, there is a high probability for strong residual self-interference. If the  amplitude of the transmit symbol of the relay exceeds the threshold $ x_{\rm th}$, the chance for very strong residual self-interference becomes too high, and the source remains silent to conserve energy for the cases when the  residual self-interference is weaker. On the other hand, from the relay's perspective,   the relay transmits   high-amplitude symbols, i.e., symbols which have an amplitude which exceeds the threshold $x_{\rm th}$, only when the source is silent as such high amplitude symbols cause strong residual self-interference.

\subsection{Optimal Input Distribution at the Relay ${p^*(x_R)}$}

The optimal input distribution at the relay, denoted by $p^*(x_R)$, which achieves the capacity of the considered relay channel is given in the following theorem.

\begin{theorem}\label{theo_2}
 If condition
\begin{align}
 \log_2\left(1+\frac{  P_R}{\sigma_D^2}\right) 
  \leq \displaystyle\int\limits_{-x_{\rm th}}^{x_{\rm th}}  \log_2\left(1+ \frac{\alpha(x_{\rm th}^2-x_R^2)}{\sigma_R^2+\alpha x_R^2}\right) \frac{1}{\sqrt{2\pi   P_R}} e^{-\frac{x_R^2}{2  P_R}} dx_R  \label{39}
\end{align}
holds, where the amplitude threshold  $x_{\rm th}$ is found from
\begin{align}
 \sqrt{\frac{2 P_R }{\pi}} \alpha x_{\rm th}  \exp\left(-\frac{x_{\rm th}^2}{2 P_R }\right) +\alpha(x_{\rm th}^2 - P_R)\mathrm{erf}\left(\frac{x_{\rm th}}{\sqrt{2  P_R}}\right)=P_S,\label{40}
\end{align}
with $\mathrm{erf}(x)=\frac{2}{\sqrt{\pi}}\int_{0}^{x}e^{-t^2}dt$, then the optimal input distribution  at the relay, $p^*(x_R)$, is the zero-mean Gaussian distribution with variance $  P_R$ and the corresponding capacity of the considered relay channel is given by
\begin{align}
C=\frac{1}{2}\log_2\left(1+\frac{  P_R}{\sigma_D^2}\right). \label{cap_1}
\end{align}
Otherwise, if condition (\ref{39}) does not hold, then the optimal input distribution  at the relay, $p^*(x_R)$, is discrete and symmetric with respect to $x_R=0$. Furthermore, the capacity and the optimal input distribution  at the relay, $p^*(x_R)$, can be found by solving the following concave optimization problem
\begin{align} 
C=\max_{p(x_R)\in\mathcal{P}} &   \sum_{x_R\in \mathcal{X}_R} \frac{1}{2} \log_2\left(1+\frac{\alpha \max\{0,x_{\rm th}^2- x_R^2\}}{\sigma_R^2+\alpha x_R^2}\right) p(x_R)\nonumber\\
\textrm{Subject to}  \textrm{ C1: }& \sum_{x_R\in \mathcal{X}_R} \frac{1}{2} \log_2\left(1+\frac{\alpha \max\{0,x_{\rm th}^2- x_R^2\}}{\sigma_R^2+\alpha x_R^2}\right) p(x_R)  \leq I(X_R;Y_D)\nonumber\\
\textrm{ C2: }& \sum_{x_R\in \mathcal{X}_R} x_R^2 p(x_R)\leq  P_R\nonumber\\
\textrm{ C3: }& \sum_{x_R\in \mathcal{X}_R} \alpha \max\{0,x_{\rm th}^2 -  x_R^2\} p(x_R) =   P_S.
\label{cap_2}
\end{align}
Moreover, solving  (\ref{cap_2})  reveals that constraint  C1 has to hold with equality and  that $p^*(x_R)$ has  the following discrete form
\begin{align}
p^*(x_R)=p_{R,0}\delta(x_R)+\sum_{j=1}^{J}\frac{1}{2}p_{R,j}(\delta(x_R-x_{R,j})+\delta(x_R+x_{R,j})),\label{n10}
	\end{align} 
	where $p_{R,j}\in[0,1]$ is the probability that $X_R=x_{R,j}$ will occur, where $x_{R,j}>0$  and $\sum_{j=0}^{J}p_{R,j}=1$   hold. With $p^*(x_R)$ as in (\ref{n10}), the capacity has the following general form
\begin{align}\label{cap_2a}
C=    \frac{p_{R,0}}{2}   \log_2\left(1+\frac{\alpha x_{\rm th}^2}{\sigma_R^2} \right)  +  \sum_{j=1}^{J}  \frac{p_{R,j}}{2}   \log_2\left(1+\frac{\alpha \max\{0,x_{\rm th}^2- x_{R,j}^2\}}{\sigma_R^2+\alpha x_{R,j}^2}\right) .
\end{align}
\end{theorem}

\begin{IEEEproof}
	Please refer to Appendix B.
\end{IEEEproof}

From Theorem~\ref{theo_2}, we can draw the following conclusions. If condition (\ref{39}) holds, then the relay-destination channel is the bottleneck link. In particular, even if the relay transmits with a zero-mean Gaussian distribution, which   achieves the capacity of the relay-destination channel, the capacity   of the relay-destination channel is still  smaller than the  mutual information (i.e.,  data rate) of the source-relay channel. Otherwise, if condition (\ref{39}) does not hold, then the optimal input distribution at the relay, $p^*(x_R)$, is always discrete and symmetric with respect to $x_R=0$. Moreover, in this case, the   mutual informations of the source-relay and relay-destination channels have to be equal, i.e., $I(X_S;Y_R|X_R)\big|_{p(x_R)=p^*(x_R)}=I(X_R;Y_D)\big|_{p(x_R)=p^*(x_R)}$ has to hold. In addition, we note that constraint C2 in  (\ref{cap_2}) does not always have to hold with equality, i.e., in certain cases it is optimal for the relay to reduce its average transmit power. In particular, if the relay-destination  channel is very strong compared to the source-relay channel, then, by reducing the  average transmit power of the relay, we reduce the average  power of the residual self-interference in the source-relay channel, and thereby improve the quality of the source-relay channel. We note that this phenomenon was first observed in \cite{5089955} and \cite{5961159}, where it was shown that, in certain cases, it is  beneficial for FD relays  to not transmit with the maximum available average power.
However, even if the   average transmit power of the relay is reduced, $I(X_S;Y_R|X_R)\big|_{p(x_R)=p^*(x_R)}=I(X_R;Y_D)\big|_{p(x_R)=p^*(x_R)}$ still  has to hold, for the data rates of the source-relay  and the relay-destination channels to be equal. 

\begin{remark}
From (\ref{40}),  it can be observed that  threshold $x_{\rm th}$ is inversely proportional to the normalized self-interference amplification factor $\alpha$. In other words, the smaller $\alpha$ is, the larger   $x_{\rm th}$ becomes.  In the limit, when  $\alpha\to 0$, we have $x_{\rm th}\to\infty$.  This is expected since for smaller  $\alpha$, the average power of the residual self-interference also becomes smaller,  which allows the source to  transmit more frequently. If  $\alpha\to 0$  the residual self-interference   tends to zero. Consequently,  the source should never be silent, i.e., $x_{\rm th}\to\infty$, which is in line with the optimal behavior  of the source for the case of ideal FD relaying without residual self-interference described in \cite{cover}.
On the other hand,   inserting the solution for $x_{\rm th}$ from (\ref{40}) into (\ref{39}), and then evaluating  (\ref{39}), it can be observed that the right hand-side of (\ref{39}) is a strictly decreasing function of $\alpha$. This is expected  since larger $\alpha$   result  in a residual self-interference with larger average  power   and thereby in a smaller achievable rate on the source-relay channel.  
\end{remark}


\subsection{Achievability of the Capacity}

The source wants to transmit  message $W$ to the destination, which is drawn uniformly from a message set $\{1,2,...,2^{nR}\}$  and   carries $nR$ bits of information, where $n\to\infty$. To this end, the transmission time is split into $B+1$ time slots and each time slot is comprised of $k$ symbol intervals, where $B\to\infty$ and $k\to\infty$. Moreover,   message $W$ is split into $B$ messages, denoted by $w(1),...,w(B)$, where each $w(b)$, for $b=1,...,B$, carries $kR$ bits of information.  Each of these messages is to  be transmitted in a different time slot. In particular, in  time slot one,   the source sends message  $w(1)$   during $k$ symbol intervals to the relay and the relay is silent.  In  time slot $b$, for $b=2,...,B$, source and relay send  messages $w(b)$ and $w(b-1)$ to relay and destination  during $k$ symbol intervals, respectively. In time slot $B+1$, the relay sends  message  $w(B)$   to the destination  during $k$ symbol intervals and the source is silent. Hence,     in the first time slot, the relay is silent since it does not have information to transmit, and in time slot $B+1$,  the source is silent since it  has no more information to transmit. In time slots $2$ to $B$, both   source and relay transmit.  During the $B+1$ time slots, the channel is used $k(B+1)$ times to send $nR=BkR$ bits of information, leading to an   overall information rate   of
\begin{eqnarray}
  \lim_{B\to\infty} \lim_{k\to\infty}  \frac{ Bk R}{k(B+1)}=R \;\;\textrm{ bits/symbol}.
\end{eqnarray}

A detailed description of the proposed coding scheme for each time slot is given in  the following, where we explain the  rates, codebooks, encoding, and decoding  used for  transmission. We note that the proposed achievability scheme requires  all three nodes to have full channel state information (CSI) of the source-relay and relay-destination channels  as well as knowledge of the self-interference suppression factor $1/\hat\alpha$.

\textit{Rates: } The transmission rate  of  both source and relay is denoted by $R$ and  given by 
\begin{equation}
   R=C    -\epsilon ,\label{self-interferenceeq_r_2} 
\end{equation}
 where  $C$ is given in Theorem~\ref{theo_2}  and $\epsilon>0$ is an arbitrarily small number. 

\textit{Codebooks: }
We have two codebooks, namely, the source's transmission  codebook and the relay's transmission  codebook.
The source's transmission codebook is generated by mapping each possible binary sequence comprised of $k R$ bits, where $R$ is given by (\ref{self-interferenceeq_r_2}),  to a codeword $\mathbf{  x}_{S}$   comprised of $k p_T$  symbols, where $p_T$ is the following probability
\begin{align}\label{eq_pt}
p_T={\rm Pr}\left\{|x_R|< x_{\rm th }\right\}.
\end{align}
Hence, $p_T$ is the probability that the relay will transmit a symbol with an amplitude which is smaller than the threshold $x_{\rm th}$. In other words, $p_T$ is the fraction of symbols in the relay's codeword which have an amplitude which is smaller than the threshold $x_{\rm th}$.
 The symbols in each codeword  $\mathbf{   x}_{S}$  are generated independently according to the  zero-mean \textit{unit variance} Gaussian distribution. Since in total  there are $2^{k R}$ possible binary sequences comprised of  $k R $  bits, with this mapping, we generate $2^{k R }$  codewords $\mathbf{  x}_{S}$ each comprised of $k p_T$ symbols.  These $2^{k R }$  codewords  form the source's transmission codebook, which we  denote    by $\mathcal{  C}_{S}$.

On the other hand, the  relay's transmission  codebook is generated by mapping each possible binary sequence comprised of $k R $ bits, where $R $ is given by (\ref{self-interferenceeq_r_2}),  to a  transmission codeword  $\mathbf{x}_R$   comprised of $k$  symbols.   The   symbols in each codeword  $\mathbf{x}_R$  are generated independently according to the optimal  distribution  $p^*(x_R)$  given in Theorem~\ref{theo_2}. The  $2^{k R}$  codewords $\mathbf{x}_R$   form the relay's transmission codebook denoted  by $\mathcal{C}_R$.

The two codebooks are known at all three nodes. Moreover, the power allocation policy at the source, $P_S(x_R)$, given in (\ref{P1}), is assumed to be known at   source and relay.

\begin{remark}
We note that the source's codewords, $\mathbf{x}_S$, are shorter than  the  relay's codewords, $\mathbf{x}_R,$ since the source is silent in $1-p_T$ fraction of the symbol intervals because of the expected strong interference in those symbol intervals. Since the relay transmits during the symbol intervals for which the source is silent,  its codewords are  longer than the codewords of the source.  Note that if the silent symbols of the source are taken into account  and counted as part of the source's codeword, then both codewords will have the same length.
\end{remark}

\textit{Encoding,  Transmission, and Decoding: }
In the first time slot, the source   maps $w(1)$  to the appropriate codeword $\mathbf{  x}_{S}(1)$ from its codebook $\mathcal{   C}_{S}$. Then, codeword $\mathbf{  x}_{S}(1)$ is transmitted  to the relay, where each symbol of $\mathbf{  x}_{S}(1)$ is amplified by   $\sqrt{P_S(x_R=0)}$, where $P_S(x_R)$ is given in (\ref{P1}).  On the other hand, the relay  is scheduled to always  receive and be silent (i.e., to set its transmit symbol to zero)  during the  first time slot. However,  knowing that the codeword transmitted  by the source in the first time slot, $\mathbf{ x}_{S}(1)$, is comprised of $k p_T$ symbols, the relay constructs the received codeword, denoted by $\mathbf{  y}_{R}(1)$, only from the first  $k p_T$  received symbols. 
\begin{lemma}\label{lem_1}
The    codeword $\mathbf{  x}_{S}(1)$ sent in the first time slot can be decoded successfully from the  codeword received at the relay, $\mathbf{  y}_{R}(1)$, using a typical decoder \cite{cover2012elements} since $R $ satisfies
\begin{eqnarray}\label{eq_d_1aa}
   R  < \max_{p(x_{S}|x_R=0)} I\big(X_{S}; Y_{R}| X_R=0\big) p_T =  \frac{1}{2} \log_2\left(1+\frac{\alpha  x_{\rm th}^2 }{\sigma_R^2 }\right) p_T   .
\end{eqnarray}
\end{lemma}
\begin{IEEEproof}
Please refer to  Appendix~\ref{app_5}.
\end{IEEEproof}

In time slots $b=2,...,B$, the encoding,  transmission, and decoding are performed as follows. In   time slots $b=2,...,B$, the source and the relay map $w(b)$ and $w(b-1)$ to the appropriate codewords $\mathbf{   x}_S (b)$ and $\mathbf{x}_R(b)$ from codebooks $\mathcal{  C}_{S }$ and $\mathcal{C}_{R}$, respectively. Note that the source also knows  $\mathbf{x}_R(b)$ since  $\mathbf{x}_R(b)$ was generated from $w(b-1)$ which the source transmitted in the previous (i.e., the $(b-1)$-th)   time slot.
As a result, both  source and   relay know   the symbols in $\mathbf{x}_R(b)$ and can determine whether their   amplitudes are smaller or larger than the threshold $x_{\rm th}$.
 Hence, if the amplitude of the first symbol  in  codeword $\mathbf{x}_R(b)$ is smaller than the threshold $x_{\rm th}$, then, in the first symbol interval of time slot $b$, the source transmits the first symbol from codeword  $\mathbf{  x}_{S}(b)$ amplified by $\sqrt{P_S(x_{R,1})}$, where $x_{R,1}$ is the first symbol in relay's codeword $\mathbf{x}_R(b)$ and $P_S(x_R)$ is given by (\ref{P1}).  Otherwise, if the amplitude of the first symbol  in   codeword $\mathbf{x}_R(b)$  is larger than   threshold  $x_{\rm th}$, then the source is silent.
 The same procedure  is performed   for the $j$-th symbol interval in time slot $b$, for $j=1,...,k$. In particular,  if the amplitude of the  $j$-th symbol  in  codeword $\mathbf{x}_R(b)$ is smaller than threshold $x_{\rm th}$, then in the $j$-th symbol interval of time slot $b$,  the source transmits its next untransmitted symbol from codeword  $\mathbf{ x}_{S}(b)$ amplified by $\sqrt{P_S(x_{R,j})}$, where $x_{R,j}$ is the $j$-th symbol in relay's codeword $\mathbf{x}_R(b)$. Otherwise, if the amplitude of the $j$-th symbol  in   codeword $\mathbf{x}_R(b)$ is larger than threshold $x_{\rm th}$, then for the $j$-th symbol interval of time slot $b$,  the source is silent. On the other hand, the relay transmits all symbols from $\mathbf{x}_R(b)$ while simultaneously receiving. Let $\mathbf{\hat y}_R(b)$ denote the received codeword at the relay in time slot $b$. Then, the relay discards  those symbols from the received codeword,  $\mathbf{\hat y}_R(b)$,   for which the corresponding symbols in $\mathbf{x}_R(b)$ have  amplitudes which exceed threshold $x_{\rm th}$, and only collects the symbols in $\mathbf{\hat y}_R(b)$ for which the corresponding symbols in    $\mathbf{x}_R(b)$ have  amplitudes which are smaller than $x_{\rm th}$. The  symbols collected from $\mathbf{\hat y}_R(b)$ constitute the relay's   information-carrying received   codeword, denoted by $\mathbf{  y}_{R}(b)$, which is used for decoding.

\begin{lemma}\label{lema_2}
The codewords $\mathbf{  x}_{S}(b)$ sent in time slots $b=2,\dots,B$ can be decoded successfully at the relay from the corresponding received codewords $\mathbf{  y}_{R}(b)$, respectively,  using a jointly typical decoder  since   $R$ satisfies
\vspace*{-1mm}
\begin{align}
     R <\sum_{ x_R\in\mathcal{X}_R }\max_{p(x_{S}|x_R)} I\big(X_{S}; Y_{R}| X_R=x_R\big) p^*(x_R) = \sum\limits_{ x_R\in\mathcal{X}_R } \frac{1}{2} \log_2\left(1+\frac{\alpha \max\{0,\;x_{\rm th}^2- x_R^2\}}{\sigma_R^2+\alpha x_R^2}\right) p^*(x_R)    .   \label{self-interferenceeq_r_2b}
\end{align}
\end{lemma}
\begin{IEEEproof}
Please refer to  Appendix~\ref{app_4}.
\end{IEEEproof}

On  the other hand, the destination   listens during the entire time slot $b$ and   receives a codeword denoted by  $\mathbf{y}_D(b)$.   By    following  the ``standard" method for analyzing the probability of error for rates smaller than the capacity, given in  \cite[Sec.~7.7]{cover2012elements}, it can be shown in a straightforward manner that the destination  can successfully decode  $\mathbf{x}_R(b)$ from the received codeword $\mathbf{y}_D(b)$, and thereby obtain $w(b-1)$, since   rate $R$ satisfies  
\begin{eqnarray}
     R <I(X_R; Y_{D})\big|_{p(x_R)=p^*(x_R)},   \label{self-interferenceeq_r_2a}
\end{eqnarray}
where $I(X_R; Y_{D})$ is given in Theorem~\ref{theo_2}.

In the last (i.e., the $(B+1)$-th) time slot, the source is silent and the relay transmits $w(B)$ by mapping it to the corresponding   codeword   $\mathbf{x}_R(B+1)$ from codebook $\mathcal{C}_{R}$. The relay transmits all   symbols in codeword   $\mathbf{x}_R(B+1)$ to the destination during time slot $B+1$. The destination  can decode the received codeword in time slot $  B+1 $ successfully, since (\ref{self-interferenceeq_r_2a}) holds.

Finally, since both  relay and destination can decode their respective codewords  in each time slot, the entire message $W$ can be decoded successfully at the destination at the end of the $(B+1)$-th time slot.

\begin{figure*}
\centering
 \resizebox{1\linewidth}{!}{
\pstool[width=1\linewidth]{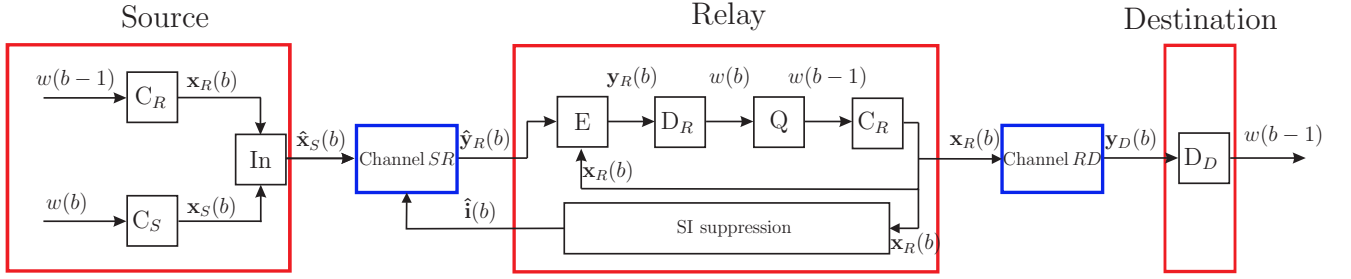}{
\psfrag{w1}[c][c][0.7]{$w(b-1)$}
\psfrag{w2}[c][c][0.7]{$w(b)$}
\psfrag{u1}[c][c][0.7]{$\mathbf{u}(b)$}
\psfrag{xs}[c][c][0.7]{$\mathbf{ x}_{S}(b)$}
\psfrag{x1}[c][c][0.7]{$\mathbf{\hat x}_S(b)$}
\psfrag{y1}[c][c][0.7]{$\mathbf{ \hat y}_R(b)$}
\psfrag{yr}[c][c][0.7]{$\mathbf{  y}_{R}(b)$}
\psfrag{x2}[c][c][0.7]{$\;\mathbf{x}_R(b)$}
\psfrag{y2}[c][c][0.7]{$\mathbf{y}_D(b)$}
\psfrag{zi}[c][c][0.7]{$\mathbf{\hat i}(b)$}
\psfrag{1}[c][c][0.8]{$\mathrm{  C}_{S}$}
\psfrag{2}[c][c][0.8]{$\mathrm{C}_R$}
\psfrag{U}[c][c][0.8]{$\mathrm{C}_R$}
\psfrag{M}[c][c][0.8]{$\mathrm{In}$}
\psfrag{E}[c][c][0.8]{$\mathrm{E}$}
\psfrag{3}[c][c][0.8]{$\mathrm{D}_R$}
\psfrag{4}[c][c][0.8]{$\mathrm{D}_D$}
\psfrag{B}[c][c][0.8]{$\mathrm{Q}$}
\psfrag{C1}[c][c][0.6]{$\mathrm{Channel}\,SR$}
\psfrag{K}[c][c][0.6]{$\mathrm{SI~suppression}$}
\psfrag{C2}[c][c][0.6]{$\mathrm{Channel}\,RD$}
\psfrag{Source}[c][c][1]{$\mathrm{Source}$}
\psfrag{Relay}[c][c][1]{$\mathrm{Relay}$}
\psfrag{Des}[c][c][1]{$\mathrm{Destination}$}
}}
\caption{ Block diagram of the proposed channel coding protocol for time slot $b$. The following notations are used in the  block diagram: $\mathrm{C}_{S}$  and $\mathrm{C}_S$ are encoders, $\mathrm{D}_R$ and $\mathrm{D}_D$ are decoders,  $\mathrm{In}$ is an inserter,  $\mathrm{E}$ is a extractor,  $\mathrm{Q}$ is a buffer, $\mathbf{\hat i}(b)$ is the residual self-interference (SI) vector in time slot $b$,  and $w(b)$ denotes the message   transmitted by the source in time slot $b$.}
\label{Fig:Channel}
 \vspace*{-3mm}
\end{figure*}

A block diagram of the proposed coding scheme  is presented in Fig.~\ref{Fig:Channel}. In particular,
in Fig.~\ref{Fig:Channel}, we show schematically the encoding,  transmission, and decoding at  source, relay, and destination. The flow of encoding/decoding in Fig.~\ref{Fig:Channel}    is as follows. Messages $w(b-1)$ and $w(b)$ are encoded into $\mathbf{x}_R(b)$ and $ \mathbf{ x}_{S}(b)$  at the source using the encoders  $\mathrm{C}_R$ and $\mathrm{ C}_{S}$,  respectively. Then, an inserter $\mathrm{In}$  is used to create a vector $\mathbf{\hat x}_{S}(b)$ by inserting the symbols of $\mathbf{  x}_{S}(b)$ into the positions of $\mathbf{\hat x}_{S}(b)$ for  which the corresponding elements of $\mathbf{x}_R(b)$ have amplitudes smaller than $x_{\rm th}$  and setting all other symbols in $\mathbf{\hat x}_{S}(b)$ to zero. Hence, vector $\mathbf{\hat x}_{S}(b)$ is identical to codeword $\mathbf{  x}_{S}(b)$ except for the added silent (i.e., zero) symbols generated at the source.
 The source then transmits $\mathbf{\hat x}_{S}(b)$ and  the relay receives the corresponding codeword $\mathbf{\hat y}_{R}(b)$.  Simultaneously, the relay   encodes  $w(b-1)$ into  $\mathbf{x}_{R}(b)$ using encoder  $\mathrm{C}_R$ and transmits it to the destination, which receives codeword $\mathbf{y}_{D}(b)$. Next, using $\mathbf{x}_R(b)$, the relay constructs $\mathbf{ y}_{R}(b)$ from $\mathbf{\hat y}_{R}(b)$ by selecting only those symbols from $\mathbf{\hat y}_{R}(b)$ for which the corresponding symbols in  $\mathbf{x}_R(b)$   have amplitudes smaller than $x_{\rm th}$. Using decoder $\mathrm{D}_R$,  the relay then decodes $\mathbf{ y}_{R}(b)$ into $w(b)$ and stores the decoded bits in its buffer $\mathrm{Q}$. On the other hand,   the destination  decodes $\mathbf{y}_{D}(b)$   into $w(b-1)$ using decoder $\mathrm{D}_D$.

\subsection{Analytical Expression for Tight Lower Bound on the Capacity}\label{sec_num_1}
 
For the non-trivial case when condition  (\ref{39}) does not hold, i.e., the relay-destination link is not the bottleneck, the capacity of the Gaussian two-hop FD relay channel with residual  self-interference is given in the form of an optimization problem, cf. (\ref{cap_2}), which is not suitable for    analysis. As a result, in this subsection, we propose a suboptimal input distribution at the relay, which yields an analytical expression for a lower bound on the capacity, derived in Theorem~\ref{theo_2}. Our  numerical results show that this lower bound is tight, at least for the considered numerical examples,  cf. Fig.~\ref{fig_1_new}. In particular, we propose that the relay  uses  the following input distribution
\begin{align}\label{eq_r_in_app}
p(x_R)=p_{\rm B}(x_R)=q \frac{1}{\sqrt{2\pi p_R/q}}\exp\left(-\frac{x_R^2}{2p_R/q}\right) + (1-q) \delta(x_R),
\end{align}
where the value of $p_R$ is optimized in the range $p_R\leq P_R$ in order for the rate to be maximized.
Hence, with probability $q$, the relay transmits a symbol from a zero-mean Gaussian distribution with variance $p_R/q$, and is silent with probability $1-q$. Since the relay transmits only in $q$ fraction of the time, the average transmit power when the relay transmits is set to $p_R/q$ in order for the average transmit power during the entire transmission time to be $ p_R$.
Now, with the input distribution $p_{\rm B}(x_R)$ in  (\ref{eq_r_in_app}),   we obtain the mutual information of the source-relay channel as
\begin{align}\label{eq_Isr-app}
  & \max_{p(x_S|x_R)\in\mathcal{P}}    I(X_S;Y_R|X_R) \bigg|_{p(x_R)=p_{\rm B}(x_R)}\\
&= q 
\int\limits_{-x_{\rm th}}^{x_{\rm th}} \hspace{-2mm} \frac{1}{2} \log_2\left(1+ \frac{\alpha(x_{\rm th}^2-x_R^2)}{\sigma_R^2+\alpha x_R^2}\right) \frac{1}{\sqrt{2\pi p_R/q}}\exp\left(-\frac{x_R^2}{2p_R/q}\right) dx_R   +  (1-q)    \frac{1}{2} \log_2\left(1+ \frac{\alpha x_{\rm th}^2}{\sigma_R^2}\right), \nonumber
\end{align}
and the mutual information of the relay-destination channel as
\begin{align}\label{eq_Ird-app}
&I(X_R;Y_D) \bigg|_{p(x_R)=p_{\rm B}(x_R)} \nonumber\\
&=- \int_{-\infty}^\infty \Bigg[   q
 \frac{1}{\sqrt{2\pi (p_R/q+\sigma_D^2)}}  \exp\left( -\frac{y_D^2}{2 (p_R/q+\sigma_D^2)} \right) +(1-q) \frac{1}{\sqrt{2\pi \sigma_D^2}}  \exp\left( -\frac{y_D^2}{2 \sigma_D^2} \right) \Bigg] \nonumber\\
&\quad\times \log_2\Bigg( q
 \frac{1}{\sqrt{2\pi (p_R/q+\sigma_D^2)}}  \exp\left( -\frac{y_D^2}{2 (p_R/q+\sigma_D^2)} \right) +(1-q) \frac{1}{\sqrt{2\pi \sigma_D^2}}  \exp\left( -\frac{y_D^2}{2 \sigma_D^2} \right) \Bigg) \nonumber\\
&~-\frac{1}{2}\log_2(2\pi e \sigma_D^2).
\end{align}
The threshold $x_{\rm th}$ in (\ref{eq_Isr-app}) and the probability $q$ in (\ref{eq_Isr-app}) and (\ref{eq_Ird-app}) are found  from the following system of two equations
\begin{align} \label{36b-1}
\left\{
\begin{array}{ll}
&q\left(\sqrt{\frac{2   p_R/q }{\pi}} \alpha x_{\rm th}  \exp\left(-\frac{x_{\rm th}^2}{2 p_R/q }\right) +\alpha(x_{\rm th}^2 - p_R/q)\mathrm{erf}\left(\frac{x_{\rm th}}{\sqrt{2  p_R/q }}\right)\right)+ (1-q)\alpha  x_{\rm th}^2 =  P_S \\
&\max\limits_{p(x_S|x_R)\in\mathcal{P}}    I(X_S;Y_R|X_R) \bigg|_{p(x_R)=p_{\rm B}(x_R)} =I(X_R;Y_D) \bigg|_{p(x_R)=p_{B}(x_R)}. 
\end{array}
\right.
\end{align}
Thereby,  $x_{\rm th}$ and $q$ are obtained as a function of $p_R$. Now, the achievable rate with the suboptimal input distribution $p_{\rm B}(x_R)$ is found by inserting   $x_{\rm th}$  and $q$  found   from (\ref{36b-1})   into (\ref{eq_Isr-app})  or  (\ref{eq_Ird-app}), and then maximizing   (\ref{eq_Isr-app})  or  (\ref{eq_Ird-app}) with respect to $p_R$ such that $p_R\leq P_R$ holds.

\section{Numerical Evaluation}\label{Sec-Num}
In this section, we numerically evaluate the capacity of the considered two-hop FD relay channel with self-interference and compare it to several benchmark schemes. To this end, we first provide the system parameters,  introduce  benchmark schemes,  and then present the numerical results.

\subsection{System Parameters}\label{set_par}
We compute the channel gains of the   source-relay ($SR$) and relay-destination ($RD$) links using the standard path loss model 
\begin{align}\label{eq_h1}
h_{L}^2=\left(\frac{c }{f_{c} 4\pi}\right)^2 d_{L}^{-\gamma},\;\; \textrm{for }L\in\{SR,RD\},
\end{align}
  where $c$ is the speed of light, $f_c$ is the carrier frequency, $d_L$ is the distance between the transmitter and the receiver of link $L$, and $\gamma$ is the path loss exponent.   For the numerical examples  in this section, we assume $\gamma=3$, $d_{SR}=500$m, and $d_{RD}=500$m or $d_{RD}=300$m. Moreover,   we assume  a carrier frequency of  $f_c=2.4$ GHz. The  transmit bandwidth is assumed to be $200$ kHz.   Furthermore, we assume that the noise power per Hz is $-170$ dBm, which for $200$ kHz  leads to a total noise power of  $2\times 10^{-15}$ Watt. Finally, the normalized   self-interference amplification factor, $\alpha$, is computed  as $\alpha=\hat \alpha/h_{SR}^2$, where  $\hat \alpha$ is the  self-interference amplification factor. 
For our numerical results, we will assume that the   self-interference amplification factor $\hat \alpha$ ranges from $-110$ dB to $-140$ dB, hence, the self-interference suppression factor,  $1/\hat \alpha$, ranges from $110$ dB to $140$ dB. We note that self-interference suppression schemes that suppress the  self-interference by up to $110$ dB  in certain   scenarios are already available today \cite{Bharadia:2013:FDR:2486001.2486033}. Given the current research efforts and the steady advancement of technology, suppression factors of up to 140 dB in ceratin scenarios might be  possible in the near future.

\subsection{Benchmark Schemes}\label{Bench}

\textit{Benchmark Scheme 1 (Ideal FD Transmission without Residual Self-Interference):}
The idealized case is when the relay can cancel all of its  residual self-interference. For this case, the capacity of the Gaussian two-hop FD relay channel without self-interference is given in \cite{cover} as
 \begin{align}\label{eq_c_fd_ideal}
C_{\rm FD,Ideal}=\min\left\{\frac{1}{2}\log_2\left(1+\frac{P_S}{\sigma_R^2}\right) , \frac{1}{2}\log_2\left(1+\frac{P_R}{\sigma_D^2}\right)  \right \}.
\end{align}
The optimal input distributions at source and relay are zero-mean Gaussian with variances $P_S$ and $P_R$, respectively.

\textit{Benchmark Scheme 2 (Conventional FD Transmission with Self-Interference):}
The conventional FD relaying scheme for the case  when the relay suffers from residual self-interference   uses the same input distributions at  source and relay as in the ideal case when the relay does not suffer from residual self-interference, i.e., the input distributions at source and relay are zero-mean Gaussian with variances $P_S$ and $p_R$, respectively, where the relay's transmit power, $p_R$, is optimized in the range $p_R\leq P_R$ such that the achieved rate is maximized. Thereby, the achieved rate is given by 
 \begin{align}\label{eq_r_fd_conv}
R_{\rm FD,Conv}=\max_{p_R\leq P_R}\min\Bigg\{\hspace{-1mm}&\int_{-\infty}^\infty \hspace{-1mm} \frac{1}{2}\hspace{-0.5mm}\log_2\hspace{-1mm}\left(1+\frac{P_S}{\sigma_R^2+\alpha x_R^2}\right)\hspace{-1mm} \frac{e^{-x_R^2/(2 p_R)}}{\sqrt{2\pi p_R}} d x_R\; 
 ;\; \frac{1}{2}\log_2\left(1+\frac{p_R}{\sigma_D^2}\right)  \bigg \}.
\end{align}

\textit{Benchmark Scheme 3 (Optimal HD Transmission):}
 The capacity of the Gaussian two-hop HD relay channel was derived in \cite{zlatanov2014capacity-globecom}, but can also be directly obtained from Theorem~\ref{theo_2} by letting $\alpha\to\infty$. This capacity can be obtained numerically and   will be denoted   by $C_{\rm HD}$. 
 In this case, the optimal input  distribution at the relay is discrete. On the other hand, the source   transmits using a Gaussian input distribution with constant variance. Moreover, the source transmits   only when the relay is silent, i.e., only when the relay transmits  the symbol zero,  otherwise, the source is silent. Since both source and relay are silent in fractions of the time,  the average powers at source and relay for HD relaying  are adjusted such that they are equal  to the  average powers at source and relay for FD relaying, respectively.

\textit{Benchmark Scheme 4 (Conventional HD Transmission):}
The conventional HD relaying scheme   uses  zero-mean Gaussian distributions with variances $P_S$ and $P_R$ at   source and   relay, respectively. However, compared to the optimal HD transmission in \cite{zlatanov2014capacity-globecom}, in conventional HD transmission, the relay alternates between receiving and transmitting in a codeword-by-codeword manner. As a result,   the achieved rate is given by \cite{1435648}
 \begin{align}\label{eq_r_hd_conv}
R_{\rm HD,Conv}=\max_{t}\min\Bigg\{ \frac{1-t}{2}\log_2\left(1+\frac{P_S/(1-t)}{\sigma_R^2}\right);  \frac{t}{2}\log_2\left(1+\frac{P_R/t}{\sigma_D^2}\right)\Bigg\}.
\end{align}
In (\ref{eq_r_hd_conv}),  since  source and relay transmit only in $(1-t)$ and $t$ fraction of the time, the average powers at source and relay are adjusted such that they are equal to the  average powers at source and relay for FD relaying, respectively.

\begin{remark}
We note that Benchmark Schemes 1-4 employ DF relaying.
We do not consider the rate achieved with  amplified-and-forward (AF) relaying because it was shown in \cite{cover} that the optimal mode of operation for   relays in terms of rate for the class of degraded relay channels,   which the investigated two-hop relay channel belongs to, is the DF mode. This means that for the considered two-hop relay channel, the rate achieved with  AF relaying will be equal to or smaller than that achieved  with   DF relaying.
\end{remark}

\subsection{Numerical Results}\label{NumRes}

In this subsection, we denote the capacity of the considered FD relay channel, obtained from Theorem~\ref{theo_2}, by $C_{\rm FD}$.

\begin{figure}[t]
	\centering\includegraphics[width=5in]{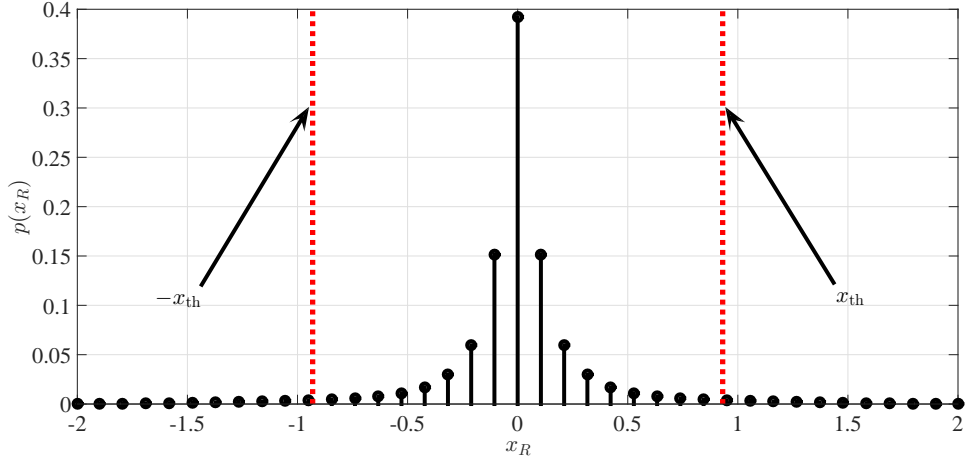}
	\caption{Optimal input distribution  at the relay, $p^*(x_R)$, for $d_{SR}= d_{RD}=500$m, $P_S=P_R=25$ dBm, and self-interference suppression factor, $1/\hat\alpha=130$ dB.}
	\label{Dis1}
\end{figure}

In Fig.~\ref{Dis1}, we plot the optimal  input distribution at the relay, $p^*(x_R)$, for $d_{SR}= d_{RD}=500$m, $P_S=P_R=25$ dBm, and a self-interference suppression factor of $1/\hat \alpha=130$  dB.   As can be seen from Fig.~\ref{Dis1}, the relay is silent in $40$\% of the time, and the source transmits only when $|x_R|<x_{\rm th}=0.9312$. Hence, similar to optimal HD relaying in \cite{zlatanov2014capacity-globecom},  shutting down the transmitter at the relay in a symbol-by-symbol manner is important for achieving the capacity. This means that in a fraction of the transmission time, the FD relay is silent and effectively works as an HD relay.
 However, in contrast to optimal HD relaying where the source transmits only when the relay is silent, i.e., only when $x_R=0$ occurs, in   FD relaying, the source has more opportunities to transmit since it can transmit also when the  relay transmits a symbol whose amplitude is smaller than $x_{\rm th}$, i.e., when $- x_{\rm th} \leq x_R\leq x_{\rm th}$ holds. For the example in Fig.~\ref{Dis1}, the source transmits  $96$~\% of the time.

 \begin{figure}
\centering
\includegraphics[width=5in]{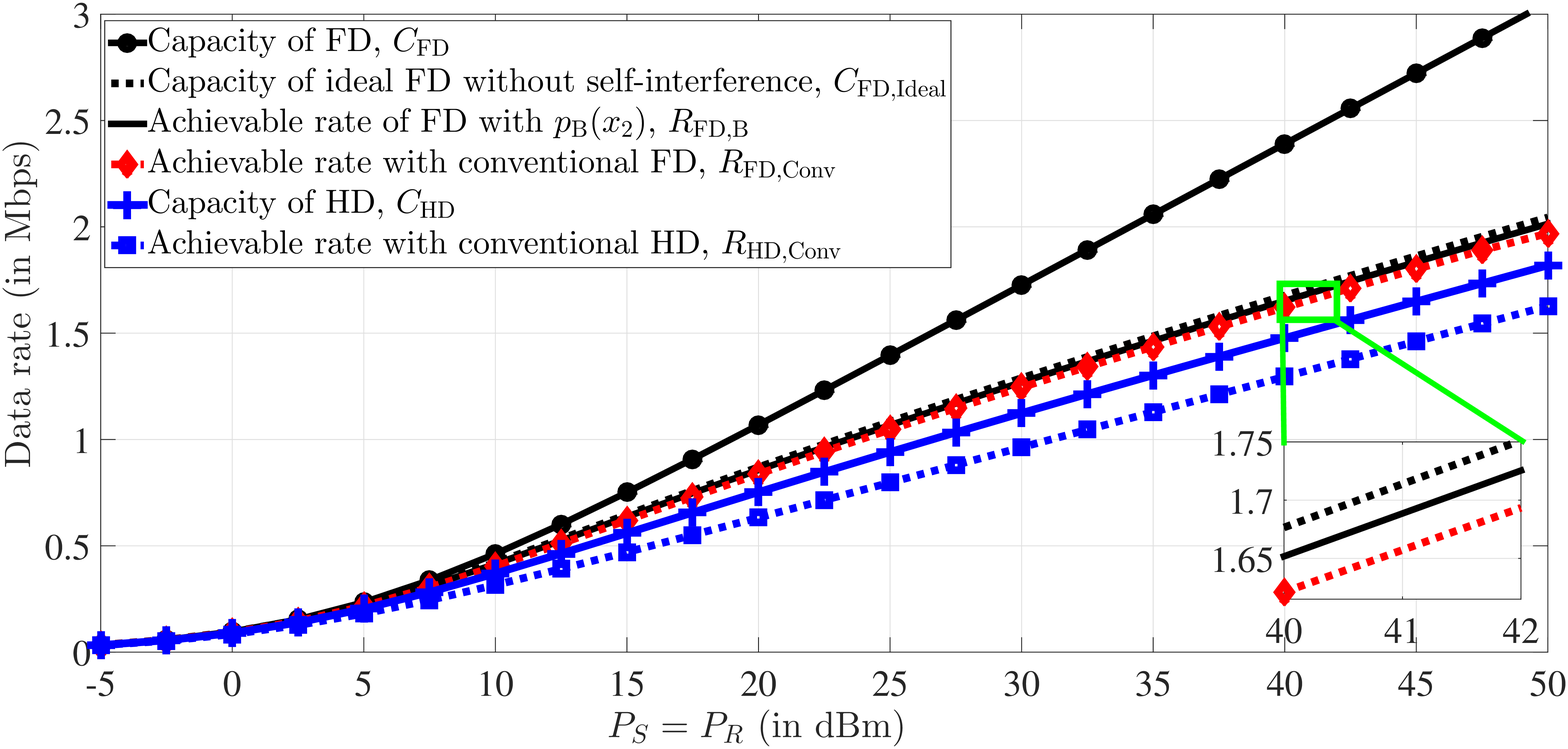}
\caption{Comparison of the derived capacity with the   rates of the benchmark schemes  as a function of the source  and relay transmit powers $P_S=P_R$ in dBm  for a self-interference suppression factor, $1/\hat\alpha= 130$ dB.}  \label{fig_1_new}
\end{figure}

  \begin{figure}
\centering
\includegraphics[width=5in]{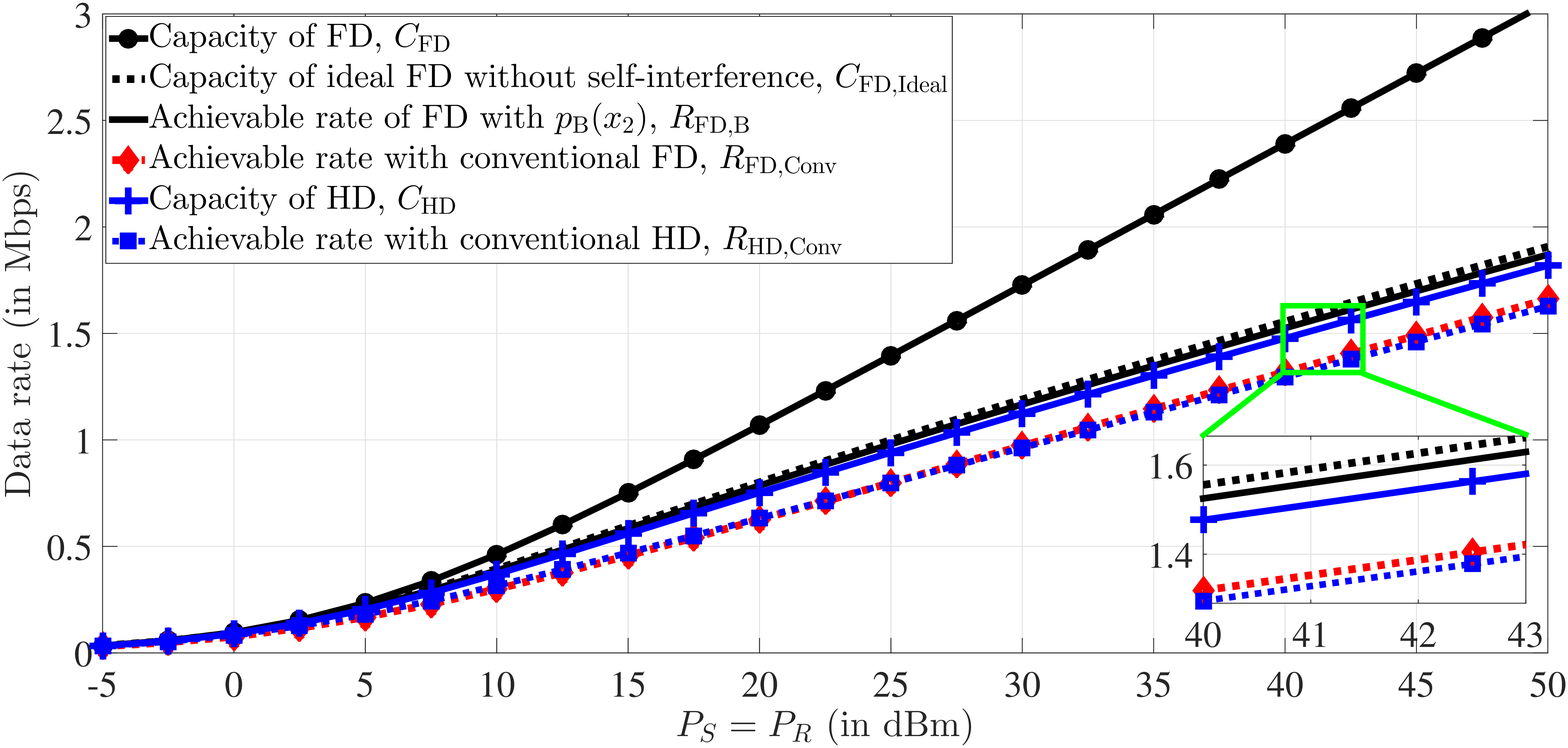}
\caption{Comparison of the derived capacity with the   rates of the benchmark schemes  as a function of the source  and relay transmit powers $P_S=P_R$ in dBm  for a self-interference suppression factor, $1/\hat\alpha= 120$ dB.}  \label{fig_2_new}
\end{figure}

 \begin{figure}
\centering
\includegraphics[width=5in]{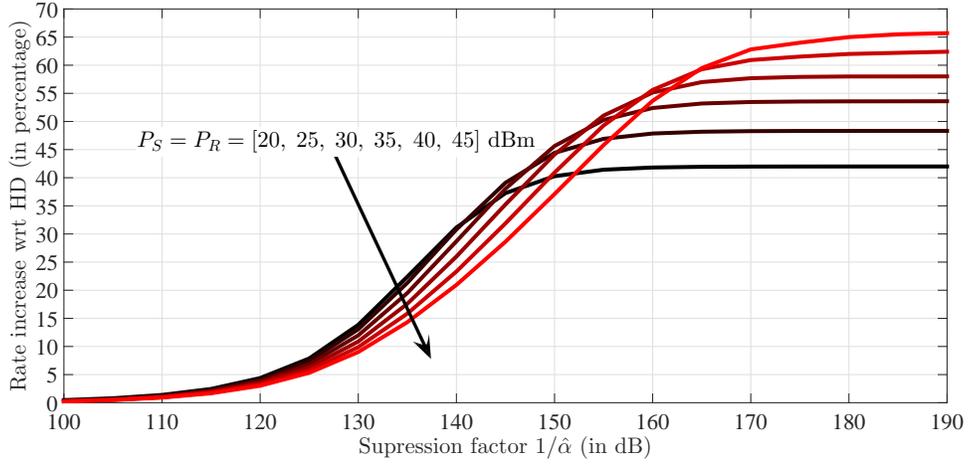}
\caption{Capacity gain of optimal FD relaying compared to optimal HD relaying as a function of the self-interference suppression factor, $1/\hat\alpha$, for different average transmit powers at source and relay $P_S$ and $P_R$.}  \label{fig_3_new}
\end{figure}

 \begin{figure}
\centering
\includegraphics[width=5in]{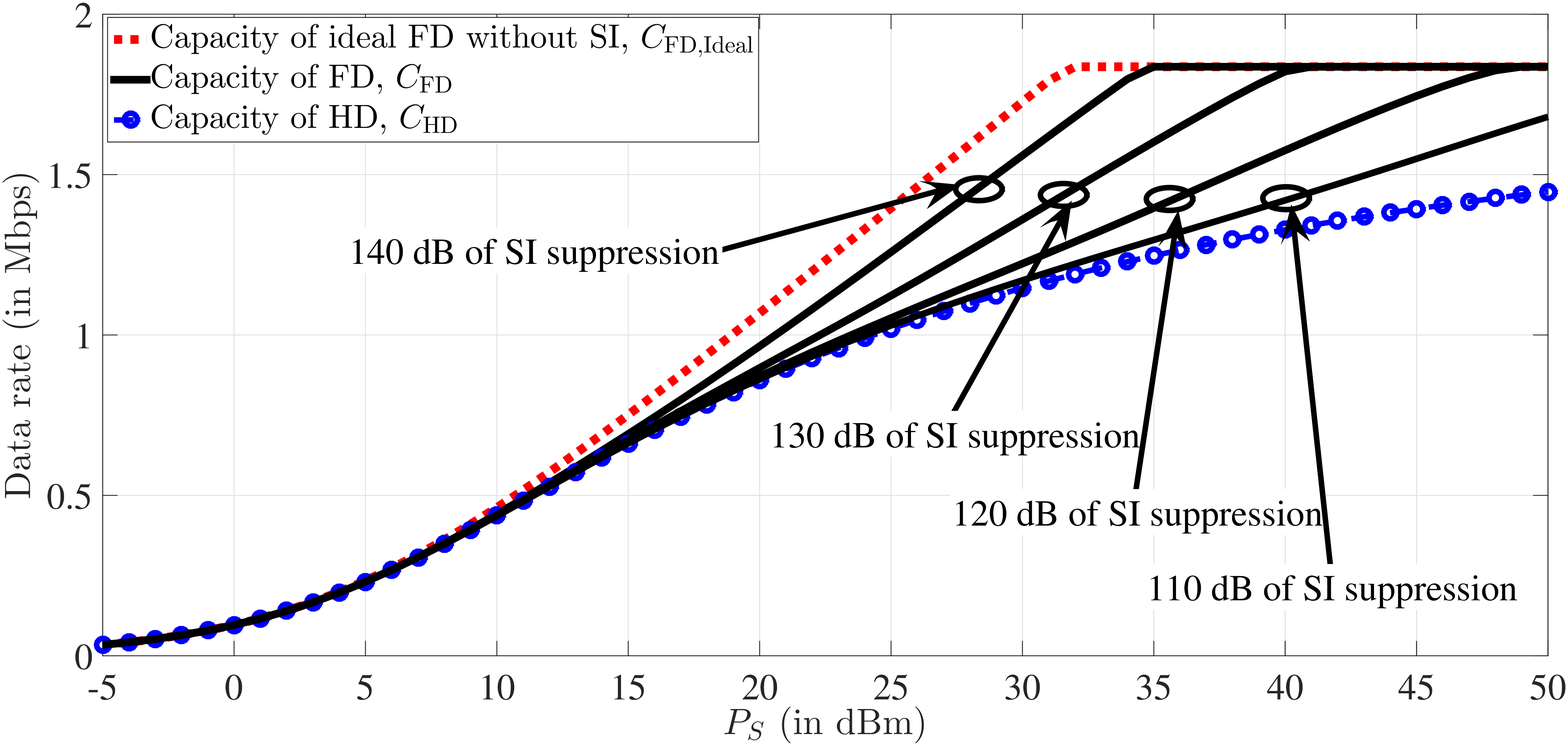}
\caption{Comparison of the derived capacity with the capacities achieved with ideal FD and optimal HD relaying as a function of the source's average transmit power $P_S$ in dBm for a fixed transmit power at the relay of $P_R=25$ dBm, and  for different self-interference (SI) suppression factors, $1/\hat \alpha$.}  \label{fig_4_new}
\end{figure}

In Fig.~\ref{fig_1_new}, we compare the capacity of the considered FD relay channel, $C_{\rm FD}$, with the achievable rate for the   suboptimal input distribution, $p_{\rm B}(x_R)$, given  in Section~\ref{sec_num_1}, denoted by $R_{\rm FD,B}$,
  the capacity achieved with ideal FD relaying without residual self-interference,  $C_{\rm FD, Ideal}$, cf. Benchmark Scheme 1,  the   rate achieved with  conventional FD relaying, $R_{\rm FD,Conv}$, cf. Benchmark Scheme 2, the capacity of the two-hop HD relay channel, $C_{\rm HD}$, cf. Benchmark Scheme 3, and the   rate achieved with  conventional HD relaying, $R_{\rm HD,Conv}$,  cf. Benchmark Scheme 4, for $d_{SR}= d_{RD}=500$m  and a self-interference suppression factor, $1/\hat\alpha$, of  130  dB as a function of the average source  and relay transmit powers $P_S=P_R$. The figure shows  that indeed the achievable rate with the   suboptimal input distribution given  in Section~\ref{sec_num_1}, $R_{\rm FD,B}$,   is a tight lower bound on the capacity $C_{\rm FD}$. Hence, this rate can be used for analytical analysis instead of the actual capacity rate, which is hard to  analyze. In addition, the figure shows  that for $P_S=P_R> 20$ dBm, the derived capacity $C_{\rm FD}$ achieves around $1.5$ dB power gain with respect to the rate achieved with conventional FD relaying, $R_{\rm FD,Conv}$, using Benchmark Scheme~2. Also, for $P_S=P_R> 20$ dBm, the the derived capacity $C_{\rm FD}$  achieves around 5 dB power gain compared to capacity of  the two-hop HD relay channel, $C_{\rm HD}$,  and around 10  dB power gain  compared to rate achieved with conventional HD relaying, $R_{\rm HD,Conv}$.

In Fig.~\ref{fig_2_new},  the same parameters   as for Fig.~\ref{fig_1_new} are adopted, except  that a self-interference suppression factor, $1/\hat\alpha$, of  120  dB is assumed. Thereby, Fig.~\ref{fig_2_new} shows that for $P_S=P_R> 20$ dBm, the derived capacity $C_{\rm FD}$ achieves around $2$ dB gain with respect to capacity of    the two-hop HD relay channel, $C_{\rm HD}$, and around 6 dB gain with respect to the rates achieved with conventional FD relaying, $R_{\rm FD,Conv}$, and conventional HD relaying, $R_{\rm HD,Conv}$. In this example, we can see that, due to the strong residual self-interference,  the rate achieved with convectional FD relaying is considerably lower than the derived capacity of FD relaying and even the capacity of HD relaying.

\begin{remark}
 From Figs.~\ref{fig_1_new} and \ref{fig_2_new}, we observe that  the multiplexing gain of the derived capacity is $1/2$, i.e.,  the same as the value for the HD case. In fact,  when $P_S=P_R$, the derived  capacity for FD relaying achieves only several dB   power gain  compared to the capacity for  HD relaying. This means that for  the adopted worst-case linear residual self-interference model, a self-interference suppression factor of 130 dB is too small   to yield a multiplexing gain of 1 in the considered range of $P_S=P_R$. Intuitively, this happens since for $P_S=P_R> 15$ dBm, the average power of the residual self-interference, $\hat \alpha E\{ X_R^2[i]\}$, exceeds the average power of the Gaussian noise. In fact, in general, for a fixed self-interference   suppression factor $1/\hat \alpha$ and   $P_S=P_R\to\infty$, the power of the residual self-interference at the relay  also becomes infinite. As a result, the corresponding multiplexing gain is limited to $1/2$.
\end{remark}

In Fig.~\ref{fig_3_new}, we show the capacity gain of the two-hop FD relay channel compared to  the two-hop HD relay channel  as a function of the self-interference suppression factor, $1/\hat\alpha$, for different average transmit powers at source and relay $P_S=P_R$ and $d_{SR}= d_{RD}=500$m\footnote{In Fig. 5,  for certain values of $1/\hat \alpha$,  the capacity gain  decreases as   $P_S=P_R$ increases. This is because in this range of  $1/\hat \alpha$, the  capacity  achieved with   HD relaying   increases  faster with $P_S=P_R$ than the capacity achieved with FD relaying.}. As can be seen from Fig.~\ref{fig_3_new}, for a self-interference suppression factor of  120 dB, we obtain only a  5  percent capacity  increase for FD relaying compared to   HD relaying. In contrast,  for a self-interference suppression factor of  130 dB, we obtain around 10 to 15 percent increase in  capacity depending on the average transmit power. A  50  percent increase in  capacity is  possible if  $P_S$ and $P_R$ are larger than 25 dBm and the self-interference suppression factor is larger than  150 dB. However, such large  self-interference suppression factors might be difficult the realize in practice.

In Fig.~\ref{fig_4_new}, we compare the capacity of the considered FD relay channel, $C_{\rm FD}$, with the capacities  of the   ideal FD relay channel without self-interference,  $C_{\rm FD, Ideal}$, and   the   HD relay channel, $C_{\rm HD}$, for $d_{SR}= 500$m, $d_{RD}= 300$m, and $P_R=25$ dBm as a function of the average transmit power at the source, $P_S$. This models a practical scenario where the transmission from a source, e.g. a base station, is supported by a dedicated low-power  FD  relay.
Different self-interference suppression factors are considered for this scenario. For this example, since the relay transmit power is fixed, the capacity of the relay-destination channel is also fixed to around $1.84$ Mbps. As a result, the capacity of the considered relay channel cannot surpass $1.84$ Mbps.
In addition, it can be observed from Fig.~\ref{fig_4_new}  that the derived capacity of the considered    FD relay channel, $C_{\rm FD}$, is significantly larger than the capacity of the  HD relay channel, $C_{\rm HD}$  when the transmit power at the source is larger than 30 dBm. For example, for 1.5 Mbps, the power gains are  approximately 30 dB, 25 dB, 20 dB, and 15 dB  compared to   HD relaying  for   self-interference suppression factors of 140 dB, 130 dB,  120 dB, and 110 dB, respectively. This numerical example shows the benefits of using a dedicated low-power  FD  relay to support a high-power    base station.

\section{Conclusion}\label{con}
We studied the capacity of the Gaussian  two-hop FD relay channel with linear residual self-interference. For this channel, we considered the worst-case linear residual self-interference model, and thereby, obtained a capacity which constitutes a lower bound on the capacity for any other   linear residual self-interference model. We showed that   the capacity is achieved by   a zero-mean  Gaussian input distribution at the source whose   variance   depends on the amplitude of the  transmit symbols at the relay.  On the other hand,   the optimal input distribution at the relay is Gaussian only when the relay-destination link is the bottleneck link. Otherwise, the optimal   input distribution at the relay is discrete.  Our numerical results show that significant performance gains are achieved with the proposed capacity-achieving coding scheme  compared to the achievable rates of conventional HD    and/or FD relaying. In addition, we proposed a suboptimal input distribution at the relay, which, for the presented numerical examples, achieves rates that are close to the capacity achieved with the optimal input distribution at the relay.

\appendix 

\subsection{Proof of Theorem~\ref{Theo1}}\label{app_1}
We first assume that $p(x_R)$  is discrete. In addition, we assume  that  $p(x_S|x_R)$ is a continuous distribution, which will turn out to be a valid assumption. Now, 
from (\ref{con2b}), the corresponding maximization problem with respect to $p(x_S|x_R)$ is given by
\begin{align}\label{app_1-eq_1}
 \max\limits_{p(x_S|x_R)\in\mathcal{P}} & \sum\limits_{x_R\in\mathcal{X}_R} I(X_S;Y_R|X_R=x_R) p(x_R)\nonumber\\
\textrm{Subject to}  \textrm{ C1: }& \sum\limits_{x_R\in\mathcal{X}_R}\left[ \int_{x_S} x_S^2 p(x_S|x_R) dx_S\right] p(x_R)\leq   P_S.
\end{align}
Since $I(X_S;Y_R|X_R=x_R)$ is the mutual information of a Gaussian AWGN channel with noise power $\sigma_R^2+\alpha x_R^2$, cf.~(\ref{r1v3}),  the optimal distribution $p(x_S|x_R)$  that maximizes  $I(X_S;Y_R|X_R=x_R)$ is the zero-mean Gaussian distribution  with variance $P_S(x_R)$. The variance $P_S(x_R)$ has to satisfy constraint  C1 in (\ref{app_1-eq_1}). Hence, to find the variance $P_S(x_R)$, we first   substitute $p(x_S|x_R)$ in  (\ref{app_1-eq_1})   with the zero-mean Gaussian distribution with variance $P_S(x_R)$. Thereby, we obtain the following optimization problem
\begin{align} 
 \max_{P_S(x_R)}&~\sum_{x_R\in\mathcal{X}_R} \frac{1}{2} \log_2\left(1+\frac{P_S(x_R)}{\sigma_R^2+\alpha x_R^2}\right)p(x_R ) \nonumber\\
\textrm{Subject to}  \textrm{ C1:}& ~ \sum_{x_R\in\mathcal{X}_R} P_S(x_R) p(x_R) \leq   P_S\nonumber\\
\textrm{ C2:}& ~   P_S(x_R) \geq 0, \; \forall x_R.\label{eq_3}
\end{align}
Since (\ref{eq_3}) is a concave optimization problem, it can be solved in a straightforward manner using the Lagrangian method, which results in (\ref{P1}). In (\ref{P1}), $x_{\rm{th}}$ is a Lagrange multiplier which has to be set such that constraint C1 in (\ref{eq_3}) holds with equality. Inserting (\ref{P1}) into constraint C1 in  (\ref{eq_3}), we obtain (\ref{36a}). Whereas, inserting $P_S(x_R)$ in (\ref{P1}) into the  objective function in (\ref{eq_3}), we obtain (\ref{eq_1-dis}). 

 Following a similar procedure as above for the case when $p(x_R)$ is assumed to be continuous, we arrive at the same solution for  $P_S(x_R)$ and $\max\limits_{p(x_S|x_R)\in\mathcal{P}}   I(X_S;Y_R|X_R)$ as  in (\ref{P1})  and (\ref{eq_1-dis}), respectively, but with the sums replaced by  integrals. This concludes the proof.
 
\subsection{Proof of Theorem~\ref{theo_2}}\label{app_2}
Assuming $p(x_R)$ is discrete, the corresponding capacity expression for the $p(x_S|x_R)$ given  in Theorem~\ref{Theo1}  is given by
\begin{align} 
C=&\max_{p(x_R)\in\mathcal{P}}  ~ \mathrm{min}\left\{ \sum_{x_R\in \mathcal{X}_R}\frac{1}{2}\log_2\left(1+\frac{\alpha \max\{0,x_{\rm th}^2 - x_R^2\}}{\sigma_R^2+\alpha x_R^2}\right)p(x_R) ,I(X_R;Y_D)\right\}\nonumber\\
\textrm{Subject to}  \textrm{ C1: }& \sum_{x_R\in \mathcal{X}_R} \alpha \max\{0,x_{\rm th}^2 - x_R^2\} = P_S\nonumber\\
\textrm{ C2: }& \sum_{x_R\in \mathcal{X}_R} x_R^2 p(x_R) \leq P_R  \label{con2c}
\end{align}
Using its epigraph form, the optimization problem in (\ref{con2c}) can be  equivalently represented as
\begin{eqnarray}\label{MPR2}
\begin{array}{rl}
 {\underset{p(x_R),\; u}{\rm{Maximize: }}}& u \\
{\rm{Subject\;\; to  }}\;\;  
  {\rm C1:}&\; u- \sum\limits_{x_R\in \mathcal{X}_R}\frac{1}{2}\log_2\left(1+\frac{\alpha\max\{0,x_{\rm th}^2 - x_R^2\}}{\sigma_R^2+\alpha x_R^2}\right)p(x_R) \leq 0  \\
 {\rm C2:}&\; u- I(X_R;Y_D) \leq 0  \\
  {\rm C3:}&\;  \sum\limits_{x_R\in \mathcal{X}_R} \alpha \max\{0,x_{\rm th}^2 - x_R^2\}p(x_R) = P_S  \\
 {\rm C4:}&\;  \sum\limits_{x_R\in \mathcal{X}_R} x_R^2 p(x_R) \leq P_R \\
{\rm C5:}&\;    \sum\limits_{x_R\in \mathcal{X}_R}   p(x_R) -1=0.
\end{array}
\end{eqnarray}
In the optimization problem   (\ref{MPR2}), constraint C2 is convex with respect to $p(x_R)$, and constraints C1, C3, C4, and C5 are affine with respect to $p(x_R)$. Hence, the optimization problem in (\ref{MPR2}) is a concave optimization problem and can be solved using the Lagrangian method. The Lagrangian function of the optimization problem in  (\ref{MPR2}) is given by
\begin{align}\label{eq_1}
&L=  u-\xi_1 \left(u- \sum\limits_{x_R\in \mathcal{X}_R}\frac{1}{2}\log_2\left(1+\frac{\alpha \max\{0,x_{\rm th}^2 - x_R^2\}}{\sigma_R^2+\alpha x_R^2}\right)p(x_R) \right) - \xi_2 \left( u- I(X_R;Y_D) \right) \nonumber\\
&-\lambda_1 \left(\sum\limits_{x_R\in \mathcal{X}_R} \alpha \max\{0,x_{\rm th}^2 - x_R^2\} p(x_R) - P_S \right) - \lambda_2 \left(  \sum\limits_{x_R\in \mathcal{X}_R} x_R^2 p(x_R) - P_R  \right) -\nu \left( \sum\limits_{x_R\in \mathcal{X}_R}   p(x_R) -1 \right),
\end{align}
where $\xi_1$, $\xi_2$, $\lambda_1$, $\lambda_2$, and $\nu$ are Lagrange multipliers corresponding to constraints C1, C2, C3, C4, and C5, respectively.  Due to the KKT conditions, the following has to hold
\begin{subequations}\label{eq_2}
\begin{align}
&\xi_1 \left(u- \sum\limits_{x_R\in \mathcal{X}_R}\frac{1}{2}\log_2\left(1+\frac{\alpha \max\{0,x_{\rm th}^2 - x_R^2\}}{\sigma_R^2+\alpha x_R^2}\right)p(x_R) \right)  =0,\;\;\; \xi_1\geq 0\label{eq_2a}\\
& \xi_2 \left( u- I(X_R;Y_D) \right) =0, \;\;\; \xi_2\geq 0\label{eq_2b}\\
&\lambda_1 \left(\sum\limits_{x_R\in \mathcal{X}_R} \alpha \max\{0,x_{\rm th}^2 - x_R^2\} p(x_R) - P_S \right) =0\label{eq_2c}\\
& \lambda_2 \left(  \sum\limits_{x_R\in \mathcal{X}_R} x_R^2 p(x_R) - P_R  \right) =0,\;\;\; \lambda_2\geq 0,\label{eq_2d}\\
& \nu   \left( \sum\limits_{x_R\in \mathcal{X}_R}   p(x_R) -1 \right) =0 .\label{eq_2e}
\end{align}
\end{subequations} 
 Differentiating $L$ with respect to $u$, we obtain that $\xi_1=1-\xi_2=\xi$ has to hold, where $0\leq \xi\leq 1$. Inserting this into  (\ref{eq_1}), then differentiating with respect to $p(x_R)$, and equating the result to zero  we obtain the following 
\begin{align}\label{eq_1aa}
& \xi  \frac{1}{2}\log_2\left(1+\frac{\alpha \max\{0,x_{\rm th}^2 - x_R^2\}}{\sigma_R^2+\alpha x_R^2}\right)  + (1-\xi)   I'(X_R;Y_D)  \nonumber\\
&-\lambda_1    \alpha \max\{0,x_{\rm th}^2 - x_R^2\}  - \lambda_2   x_R^2     -\nu     =0,
\end{align}
where  $I'(X_R;Y_D)=\partial I (X_R;Y_D)/\partial p(x_R)$.
 We note that there are  three possible solutions for (\ref{eq_1aa}) depending on whether $\xi= 1$,  $\xi=0$, or $0<\xi< 1$, respectively. In the following, we analyze these three cases.

\textit{Case 1:} Let us assume that $\xi=1$ holds. Then, from (\ref{eq_2}), we obtain that 
\begin{align}\label{eq_8}
u< I(X_R;Y_D) \textrm{ and }
u =\sum\limits_{x_R\in \mathcal{X}_R}\frac{1}{2}\log_2\left(1+\frac{\alpha \max\{0,x_{\rm th}^2 - x_R^2\}}{\sigma_R^2+\alpha x_R^2}\right)p(x_R)  ,
\end{align}
which means that for the optimal $p(x_R)$  the following holds
\begin{align}\label{eq_9}
 I(X_R;Y_D)\bigg|_{p(x_R)=p^*(x_R)}>\sum\limits_{x_R\in \mathcal{X}_R}\frac{1}{2}\log_2\left(1+\frac{\alpha \max\{0,x_{\rm th}^2 - x_R^2\}}{\sigma_R^2+\alpha x_R^2}\right)p^*(x_R).
\end{align}
The optimal $p^*(x_R)$ in this case has to maximize  the right hand side of (\ref{eq_9}), i.e., 
\begin{align}\label{eq_f}
\sum\limits_{x_R\in \mathcal{X}_R}\frac{1}{2}\log_2\left(1+\frac{\alpha \max\{0,x_{\rm th}^2 - x_R^2\}}{\sigma_R^2+\alpha x_R^2}\right)p(x_R).
\end{align}
It turns out that the optimal   $p(x_R)$ which maximizes (\ref{eq_f})  is $p^*(x_R)=\delta(x_R)$, i.e., the relay is always silent and never transmits. However, if we insert $p^*(x_R)=\delta(x_R)$ in $I(X_R;Y_D)$ in (\ref{eq_9}), we obtain the following contradiction
\begin{align}\label{eq_9a}
 I(X_R;Y_D)\bigg|_{p(x_R)=\delta(x_R)} =0 > \frac{1}{2}\log_2(1+P_S/\sigma_R^2) >0 .
\end{align}
Hence, $\xi=1$ is not possible. The only remaining possibilities are $\xi=0$ and $0<\xi<1$.

\textit{Case 2:} Let us assume that $\xi=0$ holds. Then, from (\ref{eq_2}), we obtain that 
\begin{align}\label{eq_4}
u= I(X_R;Y_D) \textrm{ and }
u< \sum\limits_{x_R\in \mathcal{X}_R}\frac{1}{2}\log_2\left(1+\frac{\alpha \max\{0,x_{\rm th}^2 - x_R^2\}}{\sigma_R^2+\alpha x_R^2}\right)p(x_R)    ,
\end{align}
has to hold,
which means that for the optimal $p(x_R)$  the following holds
\begin{align}\label{eq_5}
 I(X_R;Y_D)\bigg|_{p(x_R)=p^*(x_R)}< \sum\limits_{x_R\in \mathcal{X}_R}\frac{1}{2}\log_2\left(1+\frac{\alpha \max\{0,x_{\rm th}^2 - x_R^2\}}{\sigma_R^2+\alpha x_R^2}\right)p^*(x_R)  .
\end{align}
The optimal $p(x_R)$ in this case is the one which maximizes  the left hand side of (\ref{eq_5}), i.e.,   maximizes  $I(X_R;Y_D)$. Since the relay-destination link is an AWGN channel, $I(X_R;Y_D)$ is maximized for $p^*(x_R)$ being the zero-mean Gaussian distribution with variance $P_R$. As a result, the capacity is given by
\begin{align}\label{eq_6}
I(X_R;Y_D)\bigg|_{p(x_R)=p^*(x_R)} =\frac{1}{2} \log_2\left(1+\frac{P_R}{\sigma_D^2}\right).
\end{align}
Hence, (\ref{eq_6}) is the capacity if and only if (iff) after substituting   $p^*(x_R)$ with the zero-mean Gaussian distribution with variance $P_R$,  (\ref{eq_5})   holds, i.e., (\ref{39})  holds.

\textit{Case 3:} Let us assume that $0<\xi<1$. Then, from (\ref{eq_2}), we obtain that 
\begin{align}\label{eq_10}
u= I(X_R;Y_D) \textrm{ and }
u= \sum\limits_{x_R\in \mathcal{X}_R}\frac{1}{2}\log_2\left(1+\frac{\alpha \max\{0,x_{\rm th}^2 - x_R^2\}}{\sigma_R^2+\alpha x_R^2}\right)p(x_R)  ,
\end{align}
which means that for the optimal $p(x_R)$,  the following holds
\begin{align}\label{eq_11}
 I(X_R;Y_D)\bigg|_{p(x_R)=p^*(x_R)} =\sum\limits_{x_R\in \mathcal{X}_R}\frac{1}{2}\log_2\left(1+\frac{\alpha \max\{0,x_{\rm th}^2 - x_R^2\}}{\sigma_R^2+\alpha x_R^2}\right)p(x_R).
\end{align}
For $0<\xi<1$, we can find the optimal distribution $p^*(x_R)$ as the solution of (\ref{eq_1aa}). To this end, we need to compute $I'(X_R,Y_D)$. Since 
  for the AWGN channel, $I(X_R;Y_D)=H(Y_D)-H(Y_D|X_R)$, where $H(Y_D|X_R)=\frac{1}{2}\log_2(2\pi e \sigma_D^2)$ hold, we obtain that $I'(X_R;Y_D)=H'(Y_D)$. On the other hand,   $H'(Y_D)$ for the AWGN channel is found as
\begin{align}\label{eq_13}
H'(Y_D)=-\int_{-\infty}^\infty \frac{1}{\sqrt{2\pi \sigma_D^2}} \exp\left(-\frac{(y_D-x_R)^2}{2\sigma_D^2}\right) \log_2(p(y_D)) dy_D -\frac{1}{\ln(2)}.
\end{align}
Inserting (\ref{eq_13}) into (\ref{eq_1aa}), we obtain 
\begin{align}\label{eq_14}
& \xi  \frac{1}{2}\log_2\left(1+\frac{\alpha \max\{0,x_{\rm th}^2 - x_R^2\}}{\sigma_R^2+\alpha x_R^2}\right)  - (1-\xi)  \int_{-\infty}^\infty \frac{1}{\sqrt{2\pi \sigma_D^2}} \exp\left(-\frac{(y_D-x_R)^2}{2\sigma_D^2}\right) \log_2(p(y_D)) dy_D \nonumber\\
&- (1-\xi)\frac{1}{\ln(2)}  -\lambda_1    \alpha \max\{0,x_{\rm th}^2 - x_R^2\}  - \lambda_2   x_R^2     -\nu     =0.
\end{align}
Hence, the optimal $p(x_R)$ has to produce  a $p(y_D)$ for which  (\ref{eq_14}) holds.
In Appendix~\ref{app_3}, we  prove that (\ref{eq_14}) cannot   hold if $p(x_R)$ is a continuous distribution and that  (\ref{eq_14}) can hold if $p(x_R)$ is a discrete distribution since then it has to hold only for the discrete values   $x_R\in\mathcal{X}_R$.

\begin{remark}
Although  we derived  (\ref{eq_1aa}) assuming that $p(x_R)$ was discrete, we would have  arrived at the same result if we had assumed that $p(x_R)$ was a continuous distribution. To do so, we first would have to replace the sums in the optimization problem in (\ref{MPR2}) with integrals with respect to $x_R$. Next, in order to obtain the stationary points of the  corresponding Lagrangian function, instead of the ordinary derivative, we would have to take the functional derivative and equate it to zero. This again would have led to the identity in (\ref{eq_1aa}). Hence, the conclusions drawn from  the Lagrangian and   (\ref{eq_1aa}) are also valid when  $p(x_R)$ is a  continuous distribution. 
\end{remark}

\subsection{Proof That $p(x_R)$ is Discrete When $0<\xi<1$}\label{app_3}
This proof is based on the proof for the discreteness of a distribution given in \cite{6193208}. Furthermore, similar to \cite{6193208}, to simplify the derivation of the proof, we set    $\sigma_D^2=1$. 

	First, we decompose the integral in (\ref{eq_14}) using Hermitian polynomials. To this end, we define
	\begin{align}
	\log_2(p(y_D))=\sum_{m=0}^{\infty}c_mH_m(y_D)\label{c13},
	\end{align}
	where the $c_m$, $\forall m$, are constants and $H_m(y_D)$, $\forall m$, are  Hermitian polynomials, see \cite{6193208}.
	Note that $\mathrm{ln}(p(y_D))$ is square integrable with respect to $e^{-\frac{y_D^2}{2}}$ and hence can be decomposed using a Fourier-Hermite series decomposition, see \cite{6193208}. Then, the integral in (\ref{eq_13})  with   $\sigma_D^2=1$, can be written as
	\begin{align}
	&\frac{1}{\sqrt{2\pi}}\int_{-\infty}^{\infty}e^{-\frac{(y_D-x_R)^2}{2}}\log_2(p(y_D))dy_D 
	= \frac{1}{\sqrt{2\pi}}\int_{-\infty}^{\infty}e^{\frac{y_D^2}{2}}e^{(-\frac{x_R^2}{2}+x_Ry_D)}\log_2(p(y_D))dy_D\notag\\
&\overset{(a)}{=}\frac{1}{\sqrt{2\pi}}\int_{-\infty}^{\infty}e^{\frac{y_D^2}{2}}\sum_{n=0}^{\infty}H_n(y_D)\frac{x_R^n}{n!}\sum_{m=0}^{\infty}c_mH_m(y_D)dy_D 
	\overset{(b)}{=} \sum_{m=0}^{\infty}c_m x_R^m,\label{c14}
	\end{align}
	where   $(a)$ is obtained by inserting (\ref{c13}) and using the generating function of Hermitian polynomials,  given by 
\begin{align}
e^{(-\frac{t^2}{2}+tx)}=\sum_{m=0}^{\infty}H_m(x)\frac{t^m}{m!}.\label{d2}
\end{align}
Furthermore, $(b)$ in (\ref{c14}) follows since   Hermitian polynomials are orthogonal with respect to the weight function $\omega(x)=e^{-\frac{x^2}{2}}$, i.e.,
\begin{align}
\int_{-\infty}^{\infty}H_m(x)H_n(x)\omega(x)dx=\begin{cases}
m!\sqrt{2\pi} &\textrm{if } m=n\\
0 & \textrm{otherwise}
\end{cases}\label{d1}
\end{align}
holds.
By inserting (\ref{c14}) into  (\ref{eq_14}), we obtain  
\begin{align}
 (1-\xi)   \sum_{m=0}^{\infty}c_m x_R^m & =   \xi \frac{1}{2}\log_2\left(1+\frac{\alpha \max\{0,x_{\rm th}^2 - x_R^2\}}{\sigma_R^2+\alpha x_R^2}\right)  -\lambda_1    \alpha \max\{0,x_{\rm th}^2 - x_R^2\}  - \lambda_2   x_R^2   \nonumber\\
& - (1-\xi)\frac{1}{\ln(2)}   -\nu   .\label{n1}
	\end{align}
Now, we have two cases for $|x_R|$, one when $|x_R|\geq x_{\rm th}$ and the other one when $|x_R|< x_{\rm th}$. Also, we have two cases for  $\lambda_2$,  one when $\lambda_2>0$ (constraint C2 in (\ref{con2c}) holds with equality) and the other are when $\lambda_2=0$ (constraint C2 in (\ref{con2c}) does not hold  with equality). The resulting four cases are analyzed in  the following.

\textit{Case 1:} If $|x_R|\geq x_{\rm th}$ and $\lambda_2>0$ hold, then  (\ref{n1}) simplifies to
\begin{align}
   \sum_{m=0}^{\infty}c_m x_R^m & =        -\frac{ \lambda_2}{ 1-\xi}   x_R^2     -  \frac{1}{\ln(2)}   -\frac{\nu}{1-\xi}   .\label{n1a}
	\end{align}

	Comparing the exponents in \eqref{n1a}, we obtain 
	\begin{align}\label{eq_coef_1}
	c_0 =-   \frac{1}{\ln(2)}   -\frac{\nu}{1-\xi};\quad c_1=0;\quad c_2 =\frac{\lambda_2}{1-\xi};\quad c_n =0,\quad \forall n>2.
	\end{align}
	Inserting (\ref{eq_coef_1}) into (\ref{c13}), we obtain  $p(y_D)$ as
	\begin{align}
	p(y_D)= e^{\ln(2)(c_0 H_0(y_D)+c_2 H_2(y_D))}\overset{(a)}{=}e^{\ln(2)(c_0-c_2)}e^{\ln(2) c_2y_D^2},\label{n2}
	\end{align}
where $(a)$ follows from $H_0(y_D)=1$ and $H_2(y_D)=y_D^2-1$.
This solution for $p(y_D)$ can be a valid probability density function only for $c_2<0$, which   yields a Gaussian distribution for $p(y_D)$. Now, for  $Y_D$ to be Gaussian distributed and $Y_D=X_R+N_D$  to hold, where $N_D$ is also Gaussian distributed,     $p(x_R)$ also has to be  Gaussian distributed. However, since the Gaussian distribution is unbounded in $x_R$, the Gaussian distribution $p(x_R)$ cannot hold only in the domain $|x_R|\geq x_{\rm th}$ but has to hold in the entire domain $|x_R|\leq \infty$. Hence, we have to see whether a Gaussian $p(x_R)$ is also optimal for $|x_R|<x_{\rm th}$. If we obtain that $p(x_R)$ is not Gaussian for $|x_R|<x_{\rm th}$, then $p(x_R)$  can only be discrete in the domain  $|x_R|\geq x_{\rm th}$ for any $x_{\rm th}>0$.

\textit{Case 2:} If  $|x_R|< x_{\rm th}$  and $\lambda_2>0$ hold, (\ref{n1}) simplifies to
\begin{align}
   \sum_{m=0}^{\infty}c_m x_R^m & =       \frac{ \xi}{1-\xi} \frac{1}{2}\log_2\left(1+\frac{\alpha (x_{\rm th}^2 - x_R^2)}{\sigma_R^2+\alpha x_R^2}\right)     -\frac{ \lambda_1\alpha+\lambda_2}{1-\xi}    x_R^2         -\frac{1}{1-\xi}(1/\ln(2)+\nu  +\lambda_1    \alpha  x_{\rm th}^2) .\label{n2a}
	\end{align}
We now represent the $\log_2(\cdot)$ function in  (\ref{n2a}) using a Taylor series expansion as
 \begin{align}\label{eq_ts}
\log_2\left(1+\frac{\alpha (x_{\rm th}^2 - x_R^2)}{\sigma_R^2+\alpha x_R^2}\right) = \sum_{n=0}^{\infty} (-1)^n a_n x_R^{2n}, 
\end{align}
where $a_n>0$, $\forall n$, and the  exact (positive) values of these coefficients  are not important for this proof. 
Inserting (\ref{eq_ts}) into (\ref{n2a}), we obtain
\begin{align}
   \sum_{m=0}^{\infty}c_m x_R^m & =       \frac{ \xi}{1-\xi} \frac{1}{2\ln(2)} \sum_{n=0}^{\infty} (-1)^n a_n x_R^{2n}       -\frac{ \lambda_1\alpha+\lambda_2}{1-\xi}    x_R^2         -\frac{1}{1-\xi}(1/\ln(2)+\nu  +\lambda_1    \alpha  x_{\rm th}^2) .\label{n2b}
	\end{align}
Comparing the exponents on the left hand side and the right hand side of (\ref{n2b}), we can find $c_m$ as
\begin{align}\label{eq_coef_2}
c_m=\left\{
\begin{array}{ll}
 \frac{ \xi}{1-\xi} \frac{1}{2\ln(2)}   a_0                -\frac{1}{1-\xi}(1/\ln(2)+\nu  +\lambda_1    \alpha  x_{\rm th}^2) & \textrm{ if } m=0\\
0     & \textrm{ if } m \textrm{ is odd}\\
 \frac{ \xi}{1-\xi} \frac{1}{2\ln(2)} \  (-1)  a_n x_R^{2}       -\frac{ \lambda_1\alpha+\lambda_2}{1-\xi}    x_R^2     & \textrm{ if } m=2\\   
\frac{ \xi}{1-\xi} \frac{1}{2\ln(2)}   (-1)^{m/2} a_{m/2} x_R^{m}  & \textrm{ if } m>2 \textrm{ and } m \textrm{ is even }\\ 
\end{array}
\right.
	\end{align}
Inserting (\ref{eq_coef_2}) into (\ref{c13}), we obtain  $p(y_D)$ as
	\begin{align}
	p(y_D)=e^{\ln(2)\sum_{m=0}^{\infty}c_{2m}H_{2m}(y_D) } &=e^{\ln(2)\sum_{n=0}^{\infty}q_ny_D^{2n}}=\prod_{n=0}^{\infty}e^{\ln(2) q_ny_D^{2n}},\label{c15}
	\end{align}
	where $q_n$ are known non-zero constants. Now, since $q_n>0$ for some $n\to\infty$, $p(y_D)$ in (\ref{c15}) cannot be a valid distribution since $p(y_D)$ becomes unbounded. As a result, $p(x_R)$  has to be discrete in the domain $|x_R|<x_{\rm th}$. Consequently, $p(x_R)$  also has to be  discrete in the domain $|x_R|\geq x_{\rm th}$. This concludes the proof for the case  when $\lambda_2>0$. Following a  similar procedure for $\lambda_2=0$ as for the case  when  $\lambda_2>0$, we  obtain that again  $p(x_R)$ has to be discrete in the entire domain of $x_R$.

On the other hand, $p^*(x_R)$ has to be symmetrical with respect to $x_R=0$. To prove this, assume that we have an unsymmetrical $p(x_R)$, denoted by $p_{u}(x_R)$, with only one unsymmetrical mass point  $x_{Ru}$ which has probability $p_{Ru}$. Now, let us   construct a new, symmetrical $p(x_R)$, denoted by $p_{s}(x_R)$, by making $p_{u}(x_R)$ symmetrical. In particular, in $p_{u}(x_R)$,  we first reduce the probability of the mass point  $x_{Ru}$ to  $p_{Ru}/2$. Next, we add the mass point $-x_{Ru}$ to $p_{u}(x_R)$ and set its probability to  $p_{Ru}/2$. Now, it is clear that the average power of the relay is identical for both $p_{u}(x_R)$ and $p_{s}(x_R)$. On the other hand, by making  $p(x_R)$ symmetrical, we have increased the entropy of $X_R$, i.e.,  $H(X_R)|_{p(x_R)=p_{u}(x_R)}\leq H(X_R)|_{p(x_R)=p_{s}(x_R)}$ holds. Consequently, we have increased the differential entropy of $Y_D$, i.e.,   $h(Y_D)|_{p(x_R)=p_{u}(x_R)}\leq h(Y_D)|_{p(x_R)=p_{s}(x_R)}$ holds. Now,  since for the AWGN channel $h(Y_D|X_R)$ is independent of $p(x_R)$, it follows that
 $I(X_R;Y_D)|_{p(x_R)=p_{u}(x_R)}\leq I(X_R;Y_D)|_{p(x_R)=p_{s}(x_R)}$ holds. This concludes the proof of the symmetry of $p^*(x_R)$.

\subsection{Proof of Lemma~\ref{lem_1}}\label{app_5}

Here, we only prove the non-trivial case when (\ref{39}) does not hold. The trivial case is identical to the case without self-interference and its achievability is shown \cite{cover}.

Let us assume that condition (\ref{39}) does not hold. Then, according to Theorem~\ref{Theo1}, $p(x_R)$ is discrete and the capacity $C$ is given in (\ref{cap_2}). Moreover, for the considered coding scheme, $R$ satisfies  the following
\begin{align}\label{app_3-eq_1}
R&<C=\max_{p(x_{S}|x_R)} I\big(X_{S}; Y_{R}| X_R\big) = \sum_{x_R\in\mathcal{X}_R}\max_{p(x_{S}|x_R)} I\big(X_{S}; Y_{R}| X_R=x_R\big) p^*(x_R) \nonumber\\
 &\stackrel{(a)}{=} \sum_{\substack{x_R\in\mathcal{X}_R\\|x_R|<x_{\rm th}}}\max_{p(x_{S}|x_R)} I\big(X_{S}; Y_{R}| X_R=x_R\big) p^*(x_R)   \stackrel{(b)}{\leq} \sum_{\substack{x_R\in\mathcal{X}_R\\|x_R|<x_{\rm th}}}\max_{p(x_{S}|x_R)} I\big(X_{S}; Y_{R}| X_R=0\big) p^*(x_R)  \nonumber\\
& = \max_{p(x_{S}|x_R)} I\big(X_{S}; Y_{R}| X_R=0\big) \sum_{\substack{x_R\in\mathcal{X}_R\\|x_R|<x_{\rm th}}}  p^*(x_R)  =\max_{p(x_{S}|x_R)} I\big(X_{S}; Y_{R}| X_R=0\big)  p_T \nonumber\\
& \stackrel{(c)}{=}    
  \frac{1}{2} \log_2\left(1+\frac{\alpha   x_{\rm th}^2  }{\sigma_R^2 }\right) p_T   ,
\end{align}
where $(a)$ follows since for the considered coding scheme the source is silent when $|x_R|\geq  x_{\rm th}$ and as a result $I\big(X_{S}; Y_{R}| X_R=x_R\big)=0$ for $|x_R|\geq  x_{\rm th}$,  $(b)$ follows since, for the considered relay channel,  $I\big(X_{S}; Y_{R}| X_R=x_R\big)$  is maximized for $x_R=0$, because in that case there is no residual self-interference at the relay, and $(c)$ follows from (\ref{eq_1-dis}).  

Now, note that for the considered coding scheme in time slot $1$, the source-relay channel can be seen as an AWGN channel with a fixed channel gain $\sqrt{P_S(x_R=0)}=\sqrt{\alpha}   x_{\rm th}$ and AWGN  with variance $\sigma_R^2$ which is used $k p_T$ times. Hence,  any codeword selected  uniformly from a codebook comprised of $2^{kR}$ Gaussian distributed codewords, where each codeword is comprised of $k p_T$ symbols, with $k\to\infty$ and $R$ satisfying
\begin{align}\label{eq_ss}
kR/(kp_T)< \frac{1}{2} \log_2\left(1+\frac{\alpha   x_{\rm th}^2  }{\sigma_R^2 }\right),  
\end{align}
 can be successfully decoded at the relay using a  jointly-typical decoder, see \cite{cover2012elements}. Noting that the proposed coding scheme satisfies the properties outlined above, we can conclude that the codeword transmitted in time slot 1 can be decoded successfully at the relay.

\subsection{Proof of Lemma~\ref{lema_2}}\label{app_4}

Again,  we only prove the non-trivial case when (\ref{39}) does not hold. 

In time slot $b$, for $2\leq b\leq N$, the source-relay channel can be seen equivalently as an AWGN channel with states $X_R$, where a different state $X_R=x_R$ produces a different channel gain  and a different noise variance. In particular, for channel state $X_R=x_R$, the channel gain of the equivalent AWGN channel is $\sqrt{P_S(x_R)}$ and the variance of the AWGN    is $\sigma_R^2+\alpha x_R^2$. Moreover, for this equivalent AWGN channel with states, the source has to transmit unit-variance symbols in order for the average power constraint of the original source-relay channel, given by $E\{X_S^2\}\leq P_S$, to be satisfied. Furthermore, for the equivalent AWGN channel with states, note that both source (i.e., transmitter) and relay (i.e., receiver) have  CSI  in each channel use and thereby know that the channel gain and the noise variance in channel use $j$ will be   $\sqrt{P_S(x_{R,j})}$ and $\sigma_R^2+\alpha x_{R,j}^2$, respectively. 
Now, instead of deriving a capacity-achieving coding scheme for the original source-relay channel, we can  find equivalently a capacity-achieving coding scheme for the equivalent AWGN channel\footnote{The capacity-achieving coding scheme of the original source-relay channel can be obtained straightforwardly from  the equivalent AWGN channel with states. In particular, the only modification  is that   the source has to multiply the  transmitted symbol in channel use $j$ by  $\sqrt{P_S(x_{R,j})}$.} with states. 
To this end, we will first find the capacity of an ``auxiliary AWGN channel'', using results which are already available in the literature.  Then,  we will modify the capacity-achieving coding scheme of the ``auxiliary AWGN channel'' in order to obtain a capacity-achieving coding scheme for the equivalent AWGN channel with states.

The ``auxiliary AWGN channel'' is   identical to the equivalent AWGN channel but without CSI  at the source (i.e., transmitter). The channel coding scheme that achieves the capacity of the ``auxiliary AWGN channel'' in $k\to\infty$ channel uses is  the following, see \cite{782125, 720551} for proof. The codebook is comprised of $2^{kR}$  codewords, where each codeword is comprised of $k$ symbols and each symbol is generated independently according to the zero-mean unit-variance Gaussian distribution. Moreover, the parameter   $R$ of the channel code has to  satisfy
\vspace{-3mm}
\begin{align}\label{eq__v1}
 R &< \max_{\substack{p(x'_{S}|x_R)\\E\{X_S'^2\}= 1}}  I(X_S';Y_D|X_R)\Big|_{p(x_R)=p^*(x_R)}  = \sum_{x_R\in\mathcal{X}_R}\max_{\substack{p(x'_{S}|x_R)\\E\{X_S'^2\}= 1}} I\big(X_{S}'; Y_{R}| X_R=x_R\big) p^*(x_R) \nonumber\\
&\stackrel{(a)}{=}    
  \sum\limits_{ x_R\in\mathcal{X}_R  } \frac{1}{2} \log_2\left(1+\frac{P_S(x_R)}{\sigma_R^2+\alpha x_R^2}\right) p^*(x_R)   ,\nonumber\\
&\stackrel{(b)}{=}    
  \sum\limits_{ x_R\in\mathcal{X}_R  } \frac{1}{2} \log_2\left(1+\frac{\alpha \max\{0,\;x_{\rm th}^2- x_R^2\}}{\sigma_R^2+\alpha x_R^2}\right) p^*(x_R)   ,
\end{align} 
where $X_S'$ is the input at the source   of the ``auxiliary AWGN channel'', $(a)$ follows due to the unit-variance constraint $E\{X_S'^2\}= 1$  and since for each state $X_R=x_R$ the channel is AWGN with channel gain $\sqrt{P_S(x_{R,j})}$ and noise variance $\sigma_R^2+\alpha x_{R,j}^2$,  and $(b)$ follows from (\ref{P1}).  Any codeword selected uniformly from this codebook and transmitted in $k$ channel uses can  be successfully decoded at the relay (i.e., receiver) using a  jointly typical decoder, see  \cite{782125, 720551, cover2012elements}.   
Now,    for the ``auxiliary AWGN channel''  note that the source   transmits a   symbol    during all $k$ channel uses. Hence, the source   transmits a symbol  during  channel uses for which the channel gain is zero, i.e., $\sqrt{P_S(x_R)}=0$ holds. Obviously, the  symbols transmitted when $\sqrt{P_S(x_R)}=0$  do not reach the relay  due to the zero channel gain, i.e.,  the relay  receives  only noise during these channel uses.

Now, for the  equivalent AWGN channel, we can use the same coding scheme as for the ``auxiliary AWGN channel'', but, since in this case the source   has CSI, the source can choose not to transmit during  a channel use for which  $\sqrt{P_S(x_R)}=0$ holds.  Moreover, since the source has knowledge that $\sqrt{P_S(x_{R,j})}>0$ holds in a $p_T$ fraction out of the $k$ channel uses, the source can  reduce the length of the codewords from $k$ to  $k p_T$. Thereby, the channel code for the  equivalent AWGN channel has a codebook comprised of   $2^{kR}$ Gaussian distributed codewords, where each codeword is comprised of $k p_T$ symbols. Moreover, for the  equivalent AWGN channel, the source is silent for states for which $\sqrt{P_S(x_R)}=0$ holds, i.e.,   $|x_R|\geq x_{\rm th}$ holds, and transmits a symbol from the selected codeword only when  $\sqrt{P_S(x_R)}>0$, i.e., $|x_R|< x_{\rm th}$ holds, 
which is exactly the proposed scheme. Hence,  for the proposed scheme, we can conclude that the codeword transmitted in time slot $b$, for $2\leq b\leq N$, can be decoded successfully at the relay.


\bibliography{litdab}

\begin{thebibliography}{10}
\providecommand{\url}[1]{#1}
\csname url@samestyle\endcsname
\providecommand{\newblock}{\relax}
\providecommand{\bibinfo}[2]{#2}
\providecommand{\BIBentrySTDinterwordspacing}{\spaceskip=0pt\relax}
\providecommand{\BIBentryALTinterwordstretchfactor}{4}
\providecommand{\BIBentryALTinterwordspacing}{\spaceskip=\fontdimen2\font plus
\BIBentryALTinterwordstretchfactor\fontdimen3\font minus
  \fontdimen4\font\relax}
\providecommand{\BIBforeignlanguage}[2]{{%
\expandafter\ifx\csname l@#1\endcsname\relax
\typeout{** WARNING: IEEEtran.bst: No hyphenation pattern has been}%
\typeout{** loaded for the language `#1'. Using the pattern for}%
\typeout{** the default language instead.}%
\else
\language=\csname l@#1\endcsname
\fi
#2}}
\providecommand{\BIBdecl}{\relax}
\BIBdecl

\bibitem{C_FD_SI_conf}
N.~Zlatanov, E.~Sippel, V.~Jamali, and R.~Schober, ``Capacity of the gaussian
  two-hop full-duplex relay channel with self-interference,'' in \emph{IEEE
  Globecom 2016}, Dec. 2016.

\bibitem{cover}
T.~Cover and A.~{El Gamal}, ``Capacity theorems for the relay channel,''
  \emph{IEEE Trans. Inf. Theory}, vol.~25, pp. 572--584, Sep. 1979.

\bibitem{5089955}
T.~Riihonen, S.~Werner, and R.~Wichman, ``Optimized gain control for
  single-frequency relaying with loop interference,'' \emph{IEEE Trans.
  Wireless Commun.}, vol.~8, no.~6, pp. 2801--2806, June 2009.

\bibitem{Choi:2010}
J.~I. Choi, M.~Jain, K.~Srinivasan, P.~Levis, and S.~Katti, ``Achieving single
  channel, full duplex wireless communication,'' in \emph{MobiCom '10}.\hskip
  1em plus 0.5em minus 0.4em\relax New York, NY, USA: ACM, 2010, pp. 1--12.

\bibitem{5961159}
T.~Riihonen, S.~Werner, and R.~Wichman, ``Hybrid full-duplex/half-duplex
  relaying with transmit power adaptation,'' \emph{IEEE Trans. Wireless
  Commun.}, vol.~10, no.~9, pp. 3074--3085, Sep. 2011.

\bibitem{5985554}
------, ``Mitigation of loopback self-interference in full-duplex {MIMO}
  relays,'' \emph{IEEE Trans. Signal Proces.}, vol.~59, no.~12, pp. 5983--5993,
  Dec. 2011.

\bibitem{Jain_2011}
M.~Jain, J.~Choi, T.~Kim, D.~Bharadia, S.~Seth, K.~Srinivasan, P.~Levis,
  S.~Katti, and P.~Sinha, ``Practical, real-time, full duplex wireless,'' in
  \emph{17th Annual International Conference on Mobile Computing and
  Networking}.\hskip 1em plus 0.5em minus 0.4em\relax ACM, 2011, pp. 301--312.

\bibitem{6177689}
B.~Day, A.~Margetts, D.~Bliss, and P.~Schniter, ``Full-duplex bidirectional
  {MIMO}: Achievable rates under limited dynamic range,'' \emph{IEEE Trans.
  Signal Proces.}, vol.~60, no.~7, pp. 3702--3713, Jul. 2012.

\bibitem{6280258}
B.~P. Day, A.~R. Margetts, D.~W. Bliss, and P.~Schniter, ``Full-duplex {MIMO}
  relaying: Achievable rates under limited dynamic range,'' \emph{IEEE J.
  Select. Areas Commun.}, vol.~30, no.~8, pp. 1541--1553, Sep. 2012.

\bibitem{6353396}
M.~Duarte, C.~Dick, and A.~Sabharwal, ``Experiment-driven characterization of
  full-duplex wireless systems,'' \emph{IEEE Trans. Wireless Commun.}, vol.~11,
  no.~12, pp. 4296--4307, Dec. 2012.

\bibitem{Bharadia:2013:FDR:2486001.2486033}
D.~Bharadia, E.~McMilin, and S.~Katti, ``Full duplex radios,'' in
  \emph{Proceedings of the ACM SIGCOMM 2013}.\hskip 1em plus 0.5em minus
  0.4em\relax New York, NY, USA: ACM, 2013, pp. 375--386.

\bibitem{6542771}
E.~Ahmed, A.~Eltawil, and A.~Sabharwal, ``Rate gain region and design tradeoffs
  for full-duplex wireless communications,'' \emph{IEEE Trans. Wireless
  Commun.}, vol.~12, no.~7, pp. 3556--3565, Jul. 2013.

\bibitem{6523998}
A.~Sahai, G.~Patel, C.~Dick, and A.~Sabharwal, ``On the impact of phase noise
  on active cancelation in wireless full-duplex,'' \emph{IEEE Trans. Veh.
  Technol.}, vol.~62, no.~9, pp. 4494--4510, Nov. 2013.

\bibitem{6702851}
E.~Everett, A.~Sahai, and A.~Sabharwal, ``Passive self-interference suppression
  for full-duplex infrastructure nodes,'' \emph{IEEE Trans. Wireless Commun.},
  vol.~13, no.~2, pp. 680--694, Feb. 2014.

\bibitem{6736751}
S.~Hong, J.~Brand, J.~Choi, M.~Jain, J.~Mehlman, S.~Katti, and P.~Levis,
  ``Applications of self-interference cancellation in {5G} and beyond,''
  \emph{IEEE Commun. Magazine}, vol.~52, no.~2, pp. 114--121, Feb. 2014.

\bibitem{6656015}
M.~Duarte, A.~Sabharwal, V.~Aggarwal, R.~Jana, K.~Ramakrishnan, C.~Rice, and
  N.~Shankaranarayanan, ``Design and characterization of a full-duplex
  multiantenna system for {WiFi} networks,'' \emph{IEEE Trans. Veh. Technol.},
  vol.~63, no.~3, pp. 1160--1177, Mar. 2014.

\bibitem{6782415}
D.~Korpi, T.~Riihonen, V.~Syrjala, L.~Anttila, M.~Valkama, and R.~Wichman,
  ``Full-duplex transceiver system calculations: Analysis of {ADC} and
  linearity challenges,'' \emph{IEEE Trans. Wireless Commun.}, vol.~13, no.~7,
  pp. 3821--3836, Jul. 2014.

\bibitem{6832592}
A.~Cirik, Y.~Rong, and Y.~Hua, ``Achievable rates of full-duplex {MIMO} radios
  in fast fading channels with imperfect channel estimation,'' \emph{IEEE
  Trans. Signal Proces.}, vol.~62, no.~15, pp. 3874--3886, Aug. 2014.

\bibitem{6862895}
Y.~Y. Kang, B.-J. Kwak, and J.~H. Cho, ``An optimal full-duplex af relay for
  joint analog and digital domain self-interference cancellation,'' \emph{IEEE
  Trans. Commun.}, vol.~62, no.~8, pp. 2758--2772, Aug. 2014.

\bibitem{6832471}
B.~Debaillie, D.-J. van~den Broek, C.~Lavin, B.~van Liempd, E.~Klumperink,
  C.~Palacios, J.~Craninckx, B.~Nauta, and A.~Parssinen, ``Analog/{RF}
  solutions enabling compact full-duplex radios,'' \emph{IEEE J. Select. Areas
  Commun.}, vol.~32, no.~9, pp. 1662--1673, Sep. 2014.

\bibitem{6832464}
A.~Sabharwal, P.~Schniter, D.~Guo, D.~Bliss, S.~Rangarajan, and R.~Wichman,
  ``In-band full-duplex wireless: Challenges and opportunities,'' \emph{IEEE J.
  Select. Areas Commun.}, vol.~32, pp. 1637--1652, Sep. 2014.

\bibitem{6832439}
D.~Korpi, L.~Anttila, V.~Syrjala, and M.~Valkama, ``Widely linear digital
  self-interference cancellation in direct-conversion full-duplex
  transceiver,'' \emph{IEEE J. Select. Areas Commun.}, vol.~32, no.~9, pp.
  1674--1687, Sep. 2014.

\bibitem{7105647}
M.~Heino, D.~Korpi, T.~Huusari, E.~Antonio-Rodriguez, S.~Venkatasubramanian,
  T.~Riihonen, L.~Anttila, C.~Icheln, K.~Haneda, R.~Wichman, and M.~Valkama,
  ``Recent advances in antenna design and interference cancellation algorithms
  for in-band full duplex relays,'' \emph{IEEE Commun. Magazine}, vol.~53,
  no.~5, pp. 91--101, May 2015.

\bibitem{7024120}
G.~Liu, F.~Yu, H.~Ji, V.~Leung, and X.~Li, ``In-band full-duplex relaying: A
  survey, research issues and challenges,'' \emph{IEEE Commun. Surveys
  Tutorials}, vol.~17, no.~2, pp. 500--524, Second quarter 2015.

\bibitem{7051286}
E.~Ahmed and A.~Eltawil, ``All-digital self-interference cancellation technique
  for full-duplex systems,'' \emph{IEEE Trans. Wireless Commun.}, vol.~14,
  no.~7, pp. 3519--3532, Jul. 2015.

\bibitem{7390828}
D.~Korpi, T.~Riihonen, K.~Haneda, K.~Yamamoto, and M.~Valkama, ``Achievable
  transmission rates and self-interference channel estimation in hybrid
  full-duplex/half-duplex {MIMO} relaying,'' in \emph{IEEE 82nd Vehicular
  Technology Conference (VTC Fall)}, Sept 2015, pp. 1--5.

\bibitem{7182305}
Z.~Tong and M.~Haenggi, ``Throughput analysis for full-duplex wireless networks
  with imperfect self-interference cancellation,'' \emph{IEEE Trans. Commun.},
  vol.~63, no.~11, pp. 4490--4500, Nov. 2015.

\bibitem{kramer2004models}
G.~Kramer, ``Models and theory for relay channels with receive constraints,''
  in \emph{Proc. 42nd Annual Allerton Conf. on Commun., Control, and
  Computing}, 2004, pp. 1312--1321.

\bibitem{zlatanov2014capacity-globecom}
N.~Zlatanov, V.~Jamali, and R.~Schober, ``On the capacity of the two-hop
  half-duplex relay channel,'' in \emph{Proc. of the IEEE Global Telecomm.
  Conf. (Globecom)}, San Diego, Dec. 2015.

\bibitem{1435648}
A.~Host-Madsen and J.~Zhang, ``Capacity bounds and power allocation for
  wireless relay channels,'' \emph{IEEE Trans. Inf. Theory}, vol.~51, pp. 2020
  --2040, Jun. 2005.

\bibitem{cover2012elements}
T.~M. Cover and J.~A. Thomas, \emph{Elements of Information Theory}.\hskip 1em
  plus 0.5em minus 0.4em\relax John Wiley \& Sons, 2012.

\bibitem{TSE05}
D.~Tse and P.~Viswanath, \emph{Fundamentals of Wireless Communication}.\hskip
  1em plus 0.5em minus 0.4em\relax Cambridge University Press, 2005.

\bibitem{BA-relaying-adaptive-rate}
N.~Zlatanov, R.~Schober, and P.~Popovski, ``Buffer-aided relaying with adaptive
  link selection,'' \emph{IEEE J. Select. Areas Commun.}, vol.~31, no.~8, pp.
  1530--1542, Aug. 2013.

\bibitem{841172}
M.~Medard, ``The effect upon channel capacity in wireless communications of
  perfect and imperfect knowledge of the channel,'' \emph{IEEE Trans. Inf.
  Theory}, vol.~46, no.~3, pp. 933--946, May 2000.

\bibitem{6193208}
J.~Fahs and I.~Abou-Faycal, ``Using {Hermite} bases in studying
  capacity-achieving distributions over {AWGN} channels,'' \emph{IEEE Trans.
  Inf. Theory}, vol.~58, pp. 5302--5322, Aug. 2012.

\bibitem{782125}
G.~Caire and S.~Shamai, ``On the capacity of some channels with channel state
  information,'' \emph{IEEE Trans. Inf. Theory}, vol.~45, no.~6, pp.
  2007--2019, Sep. 1999.

\bibitem{720551}
E.~Biglieri, J.~Proakis, and S.~Shamai, ``Fading channels:
  Information-theoretic and communications aspects,'' \emph{IEEE Trans. Inf.
  Theory}, vol.~44, no.~6, pp. 2619--2692, Oct 1998.

\end{thebibliography}
\bibliographystyle{IEEEtran}

\end{document}